\documentclass[10pt,letterpaper]{article}
\usepackage[top=0.85in,left=2.75in,footskip=0.75in]{geometry}

\usepackage{amsmath,amssymb}

\DeclareMathOperator{\V}{\mathbb{V}}

\usepackage{changepage}

\usepackage{textcomp,marvosym}

\usepackage{cite}
\usepackage{float}

\usepackage{nameref,hyperref}

\usepackage[right]{lineno}

\usepackage[nopatch=eqnum]{microtype}
\DisableLigatures[f]{encoding = *, family = * }

\usepackage[table]{xcolor}

\usepackage{color,soul}

\usepackage{array}

\newcolumntype{+}{!{\vrule width 2pt}}

\newlength\savedwidth

\raggedright
\usepackage[aboveskip=1pt,labelfont=bf,labelsep=period,justification=raggedright,singlelinecheck=off]{caption}

\makeatletter
\renewcommand{\@biblabel}[1]{\quad#1.}
\makeatother

\usepackage{textcomp}

\usepackage{lastpage,fancyhdr,graphicx}
\usepackage{epstopdf}
\fancyheadoffset[L]{2.25in}
\fancyfootoffset[L]{2.25in}
\usepackage{mathtools}
\usepackage{bm}
\usepackage{amsmath}\usepackage{amssymb}

\usepackage[final]{changes}

\usepackage{multirow}

\begin{document}

\vspace*{0.2in}

\begin{flushleft}
{\Large
\textbf\newline{Regional heterogeneity in left atrial stiffness impacts passive deformation in a cohort of patient-specific models} 
}
\newline
\\
Tiffany MG Baptiste\textsuperscript{1}*,
Cristobal Rodero\textsuperscript{2},
Charles P Sillett\textsuperscript{1},
Marina Strocchi\textsuperscript{2},
Christopher W Lanyon\textsuperscript{3},
Christoph M Augustin\textsuperscript{4},
Angela WC Lee\textsuperscript{2},
José Alonso Solís-Lemus\textsuperscript{2},
Caroline H. Roney\textsuperscript{5},
Daniel B. Ennis\textsuperscript{6},
Ronak Rajani\textsuperscript{1,7},
Christopher A Rinaldi\textsuperscript{1,7},
Gernot Plank\textsuperscript{4},
Richard D Wilkinson\textsuperscript{3},
Steven E Williams\textsuperscript{1},
Steven A Niederer\textsuperscript{1,2,8}
\\
\bigskip
\textbf{1} School of Biomedical Engineering and Imaging Science, King's College London, London, United Kingdom
\\
\textbf{2} National Heart and Lung Institute, Imperial College London, London, United Kingdom
\\
\textbf{3} School of Mathematical Sciences, University of Nottingham, Nottingham, United Kingdom
\\
\textbf{4} Division of Medical Physics and Biophysics, Medical University of Graz, Graz, Austria
\\
\textbf{5} School of Engineering and Materials Science, Queen Mary University of London, London, United Kingdom
\\
\textbf{6} Department of Radiology, Stanford University, Stanford, United States
\\
\textbf{7} Guy’s and St Thomas’ NHS Foundation Trust, London, United Kingdom
\\
\textbf{8} Alan Turing Institute, London, United Kingdom
\bigskip

\textcurrency Current Address: School of Biomedical Engineering and Imaging Sciences, King's College London, London, Greater London, United Kingdom 

\bigskip
* Corresponding author

Email: tiffany.baptiste@kcl.ac.uk

\end{flushleft}

\newpage
\section*{Abstract}

In atrial fibrillation (AF), atrial biomechanics are altered, reducing atrial movement. It remains unclear whether these changes are due to altered anatomy, myocardial stiffness, or constraints from surrounding structures. Understanding the causes of changed atrial deformation in AF could enhance tissue characterization and inform AF diagnosis, stratification, and treatment. We created patient-specific anatomical models of the left atrium (LA) from CT images. Passive LA biomechanics were simulated using finite deformation continuum mechanics equations. LA stiffness was represented by the Guccione material law, where $\alpha$ scaled the anisotropic stiffness parameters. Regional passive stiffness parameters were calibrated to peak regional deformation during the reservoir phase and validated against deformation transients derived from retrospective gated CT images during the reservoir and conduit phase. Physiological LA deformation varies regionally, with the roof deforming significantly less than other regions during the reservoir phase. The fitted model matched peak patient deformations globally and regionally with an average error of $0.90 \pm 0.39$ mm over our cohort. We compared deformation transients through the reservoir and conduit phases and found that the simulated deformation transients were within an average of $\pm0.38$ mm per unit time of the CT-derived deformation transients. Regional stiffness varied across the atria with average $\alpha$ values of 1.8, 1.6, 2.2, 1.6 and 2.1 across the cohort in the anterior, posterior, septum, lateral and roof regions respectively. Using mixed effect models, we found no correlation between regional patient LA deformation and regional estimates of wall thickness or regional volumes of epicardial adipose tissue. We found a significant correlation between regionally calibrated stiffness and CT-derived LA biomechanics ($p=0.023$). We have shown that regional heterogeneity in stiffness contributes to regional LA biomechanics, while anatomical features appeared less important. These findings provide insight into the underlying causes of altered LA biomechanics in AF.

\section*{Author summary}

The deformation of the left atrium (LA), or its biomechanical function, is closely linked to the health of this cardiac chamber. In atrial fibrillation (AF), atrial biomechanics are significantly altered but the underlying cause of this change is not always clear. Patient-specific models of the LA that replicate patient atrial motion can allow us to understand how factors such as atrial anatomy, myocardial stiffness and physiological constraints are linked to atrial biomechanics. We created patient-specific LA models from CT images. We fitted regional model stiffness to peak CT-derived deformation during the LA reservoir phase ($\pm0.90$ mm) and used the CT deformation transients through the reservoir and conduit phase for model validation (deformation transients fell within $\pm0.38$ mm per unit time of targets). We found that myocardial stiffness varies regionally across the LA. The regional stiffness values were significant factors contributing to regional physiological LA deformation ($p=0.023$) while features of LA anatomy, including regional wall thickness and adipose volume, were less important. 
These findings provide insight into the underlying causes of altered LA biomechanics in AF.

\nolinenumbers

\section*{Introduction}
Atrial fibrillation (AF) is the most common cardiac arrhythmia caused by the chaotic activation of the atria, leading to a rapid and irregular heart beat. AF affects 1.4 million people in the UK and represents a significant cost to the UK National Health Service \cite{Burdett2022AtrialCosts}. Patients with AF are at greater risk of heart failure, early onset-dementia and stroke. Treatment strategies for AF can be categorised as rate or rhythm control. Treatment may be delivered pharmacologically, through ablation of underlying arrhythmogenic tissue substrate or by ablation of the atrioventricular node and implanting a pacing device \cite{Brundel2022AtrialFibrillation}. Choosing the correct therapy and identifying potentially arrhythmogenic tissue remain significant challenges in treating AF patients. As such, there is a need to better understand the pathology to improve diagnosis, develop new, more targeted therapies and guide patient selection \cite{Kirchhof2017TheTherapy,Woods2014AtrialAblation}. 

The choice of approach for AF therapy is influenced by the frequency and duration of fibrillation episodes \cite{Brundel2022AtrialFibrillation}. Increased severity of AF is associated with progressive atrial remodelling, including altered myocardial tissue properties due to the development of atrial fibrosis, which itself increases tissue stiffness and arrhythmia susceptibility \cite{Walters2016ProgressionIntervention}. Fibrosis, and thus potential AF substrate, can be identified by complex fractionated atrial electrograms \cite{Nademanee2004ASubstrate}, slow conduction velocity \cite{Honarbakhsh2019StructuralFibrillation}, low voltage regions \cite{Sim2019LeftSubstrate} or regions with enhanced gadolinium contrast in cardiac magnetic resonance (MR) images \cite{Oakes2009DetectionFibrillation}. Recording atrial electrograms, conduction velocities and voltages require invasive methods, limiting their widespread application to guiding AF care  \cite{Latchamsetty2011ComplexFibrillation}. MR late enhancement imaging is challenging and results remain controversial \cite{Siebermair2017AssessmentImplications}. Thus, robust approaches to distinguish myocardial tissue properties non-invasively, with the aim of improving AF therapy are needed.

Atrial fibrosis is characterized by an increase in tissue collagen content, leading to increased local atrial passive stiffness and potentially impacting atrial biomechanics. This AF-related change in biomechanics has been demonstrated through a consistent reduction in global and regional strains in AF patients, measured using echocardiography \cite{Kuppahally2010LeftFibrillation}, cardiac MR imaging \cite{Benjamin2022AssociationAblation} and CT \cite{Sillett2024AFibrillation}. However, it is not clear if these changes in strain with AF can be attributed to changes in anatomy or tissue stiffness or result from constraints imposed by surrounding physiological structures. Determining the underlying causes that lead to changes in atrial deformation with AF would facilitate using biomechanical measures, including strain and displacement, for tissue characterization and subsequently, AF diagnosis, disease stratification and treatment. 

Healthy atrial function encompasses three phases - a reservoir phase where the atria fill, a conduit phase where the atria passively empties into the ventricles, and a booster pump phase where the atria actively contract. In AF, the atria become stiffer due to increasing fibrosis and contract less, reducing their booster function. As a result, atrial mechanics is dominated by forces induced by ventricular motion and passive atrial properties. In addition to increased fibrosis, epicardial adipose tissue (EAT) has also been associated with AF \cite{Thanassoulis2010PericardialFibrillation}. EAT contributes to the development of AF through electrical and structural remodelling of myocardium induced via direct \cite{Mahajan2015ElectrophysiologicalObesity} and indirect \cite{Haemers2017AtrialAtria} mechanisms. This remodelling might then be reflected in the observed biomechanical behaviour - in the left ventricle, increased EAT has been correlated to reduced global longitudinal strain \cite{Sun2023AssociationSyndrome}. Along with EAT, chamber wall thickness may also impact atrial biomechanics. The walls of the atria are thin but its thickness is heterogeneous. AF has been associated with localised changes in atrial wall thickness \cite{Platonov2008LeftAutopsies}. 
\added{It has been suggested that anatomical variation in atrial wall thickness may influence local mechanical behaviours, due to a demonstrated inverse relationship between wall thickness and wall stress, and consequently wall strain}
\deleted{It has been suggested that in the atria, this anatomical variation in wall thickness may play a role in determining local mechanical behaviours due to the law of Laplace where wall thickness and wall stress, and in turn wall strain, are inversely correlated} \cite{Augustin2020TheAtrium}. Myocardial stiffness, fibrosis, EAT, wall thickness and surrounding physiological constraints might each contribute to observable atrial mechanical function, individually and in combination. Therefore, accurately assessing atrial mechanics requires accounting for all these factors.

Physics and physiology constrained models provide a consistent multi-scale framework to integrate and account for anatomy, constraints imposed by surrounding structures (boundary conditions) and material properties when analysing atrial biomechanics. In these models, constitutive relationships characterise the passive stiffness of myocardium. These relationships include a number of material parameters that describe the nonlinear passive stress-strain response of myocardium and thus provide a measure of myocardial stiffness. The determination of suitable parameter values for specific patients and pathologies remains an active area of modelling research \cite{Avazmohammadi2019AMyocardium}. Historically, uni-axial and bi-axial tests performed ex-vivo on samples of ventricular  \cite{Hunter1998ModellingMuscle, Demiray1976StressesWall, Humphrey1990DeterminationEstimation} or atrial \cite{DiMartino2011InDistribution, Bellini2013MechanicalAtria} myocardium have been used for myocardial material law parameterisation. While these ex-vivo experimental measurements are useful in providing reference values for myocardial stiffness, these estimates are not patient-specific, and may not reflect the in-vivo environment, limiting their applicability \cite{Costa2001ModellingDimensions}. 

Effort has been made towards the in-vivo estimation of myocardial passive stiffness. An early method towards non-invasive, in-vivo estimation of ventricular myocardial properties was introduced in Ghista et al.  \cite{Ghista1980CardiacStates} and has since been used in \cite{Augenstein2005MethodImaging, Wang2009ModellingFunction, Xi2011AnEstimation, Genet2014DistributionTreatments, Wang2018LeftAnalysis, Palit2018InMyocardium, Marx2022RobustConfiguration, Shi2024AnMechanics}. These studies approximated the stiffness of ventricular myocardium using inverse stiffness estimation. The parameters of the myocardial constitutive relation were calibrated by matching displacements or pressure-volume curves obtained from dynamic clinical images to displacements \cite{Augenstein2005MethodImaging, Wang2009ModellingFunction, Xi2011AnEstimation, Wang2018LeftAnalysis} or volumes \cite{Genet2014DistributionTreatments, Palit2018InMyocardium, Marx2022RobustConfiguration, Shi2024AnMechanics} produced by a simulated model. Though this method of inverse estimation of patient-specific parameters has been widely accepted and implemented, the literature, \added{aside from \cite{Shi2024AnMechanics},} has only focused on in-vivo parameter estimation in the ventricles without any parameterisation in the atria. Further, previous studies, with the exception of Ghista et al. \cite{Ghista1980CardiacStates}, have not explored regional heterogeneity of myocardial stiffness parameters, despite the ability of this method to provide a local myocardial stiffness measure being highlighted as one of the benefits of the techniques when compared to the traditional global chamber stiffness obtained from the diastolic pressure-volume relationship \cite{Wang2018LeftAnalysis}. Additionally, these previous methods have not accounted for any modelling uncertainty and have had limited application to human cases.
 
In this study, we investigate the role of heterogeneity in anatomical and stiffness properties of the left atrium (LA) in dictating its passive mechanical behaviour. We focused on the LA only as AF incidence has been associated with anatomical, structural, functional and electrical remodelling of the LA \cite{Kirchhof20162016EACTS} and historically, more attention is given to the LA in attempting to better understand and treat the disease. Previous biomechanics modelling efforts targeted at AF have also focused on this chamber \cite{Feng2019AnalysisValve},\cite{Moyer2015ChangesModel}. 

To examine the role of passive stiffness in dictating LA biomechanics, we first constructed a modelling framework that reproduces physiological passive LA deformation. Using this framework, we perform patient-specific myocardial stiffness estimation in the LA by fitting simulated endocardial surface displacements of an LA model to displacement estimates from ECG-gated CT images in heart failure patients, who are expected to have remodelled atria. We hypothesised that regional variation in myocardial stiffness properties was necessary to reproduce not only the global CT-derived deformation metrics, but regional behaviour as well. Thus, we performed stiffness calibration regionally using several different Bayesian statistical techniques. \deleted{With patient-specific regional stiffness values, wall thickness measurements and estimates of EAT, we could then systematically analyse how these factors contribute to observed LA biomechanics.}

\added{LA mechanical function is complex, incorporating both active and passive phases, and exhibits regional variation across different areas of the chamber wall. This complex mechanical function is altered in AF and other highly prevalent pathologies such as heart failure with preserved ejection fraction. Factors such as LA wall thickness, surrounding EAT and the development of fibrosis, causing local stiffening in the atrial myocardium, may each impact LA biomechanics alone or in combination. The computational LA biomechanical framework presented in this manuscript allowed us to investigate each of these factors.} With this work, we make 4 main contributions to the field. Firstly, we provide a cohort of 10 patient-specific LA models that include a patient-specific wall thickness and surrounding EAT. Secondly, we present a list of parameters that are most relevant to LA reservoir function, as well as suitable ranges for each parameter. Third, \replaced{for the first time, we}{this work represents the first to} perform inverse stiffness estimation in the atria and explore regional heterogeneity in myocardial stiffness across the LA. Finally, we show that regional stiffness, and not regional wall thickness or EAT, significantly impacts deformation in the remodelled LA.

\section*{Methods}

In this section, we summarise the clinical data and LA mechanics framework used in this study. Once the modelling framework was established, we constructed a calibration pipeline that uses sensitivity analysis to find the most relevant simulation parameters. We then fit these parameters to the available clinical data using Bayesian methods. Fig \ref{fig:methods_summary} provides an overview of the methods implemented in this study.
\begin{adjustwidth}{-2.25in}{0in}
\begin{figure}[H]
    \centering
    \includegraphics[width=\linewidth]{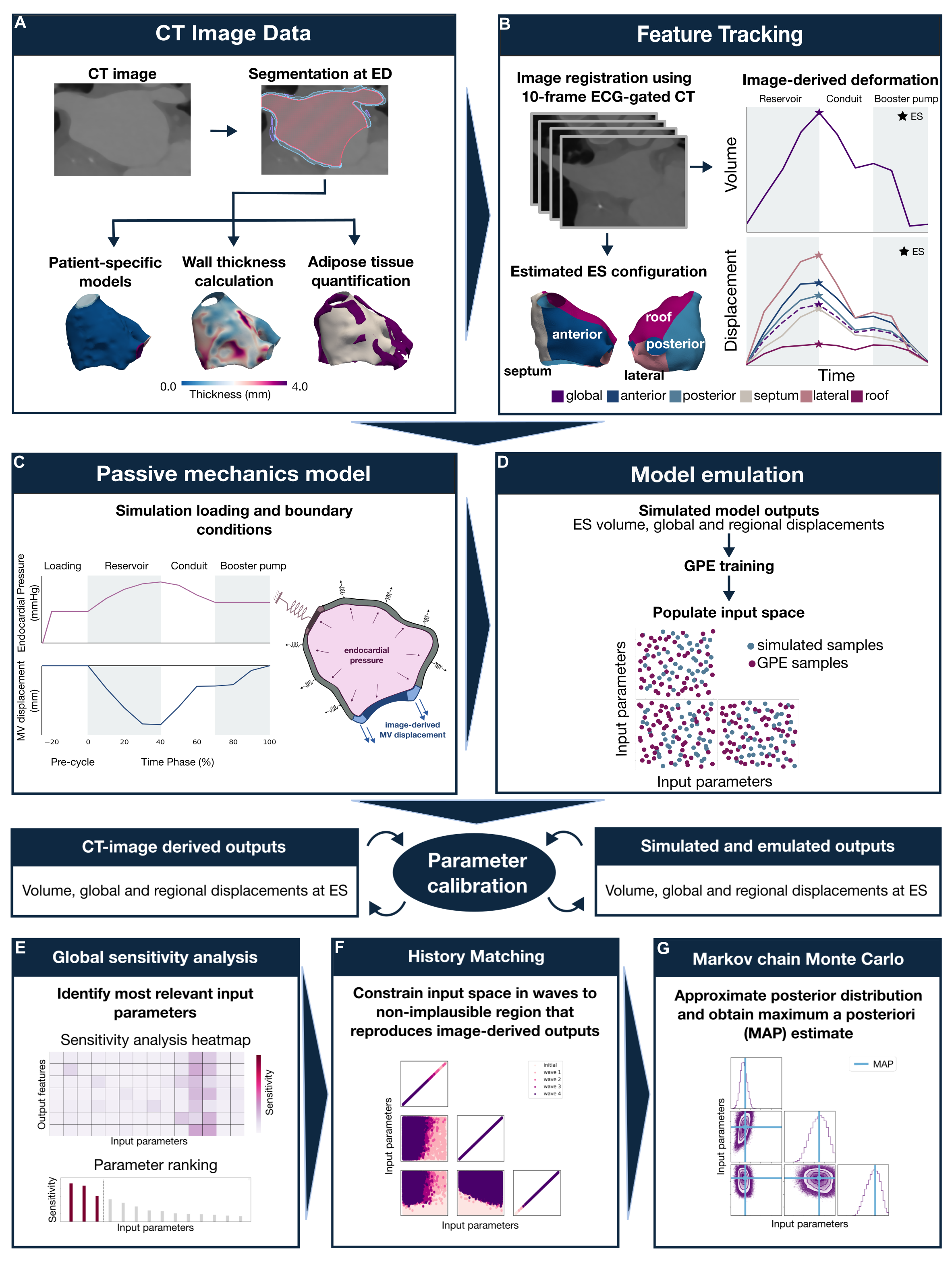}
    \caption{\textbf{Summary of Methods}. \textbf{A} A cohort of 10 LA meshes with epicardial adipose tissue (EAT) was generated from the end-diastolic (ED) frame of a gated contrast-enhanced CT image-set. \textbf{B} Using image registration, the ED LA mesh was deformed over cardiac cycle, with maximum deformation at end-systolic (ES). Plots show the endocardial volume and surface displacement transients derived from feature tracking motion models. \textbf{C} The loading conditions applied in our LA modelling framework and the simulation set-up with boundary conditions applied. A estimated pressure profile was applied to LA endocardium. Patient-specific image-derived mitral valve (MV) displacement was applied to the MV annulus in simulation model. \textbf{D} Description of how Gaussian process emulators (GPEs) were used to speed up computation time. \textbf{E-G} Summary of the fitting methods used in this study. }
    \label{fig:methods_summary}
\end{figure} 
\end{adjustwidth}

\subsection*{Clinical data}

This study complied with the Declaration of Helsinki and the protocol was approved by the West Midlands Coventry and Warwick ethics committee and the London-Harrow ethics committee (clinical trial REC numbers 14/WM/1069 and 18/LO/0752). The study was conducted in accordance with the local legislation and institutional requirements.  Each patient provided written informed consent, and images were anonymised prior to analysis. 

The atria are thin and have a complex anatomy. For this reason, the LA models were developed from retrospective ECG-gated contrast-enhanced CT acquired in 10 or 20 frames over the cardiac cycle. Due to the radiation dose associated with this imaging technique, data was collected from 10 heart failure (HF) patients with reduced ejection fraction recruited between 2014 and 2018 for cardiac resynchronisation therapy upgrade. The patients had persistent HF symptoms on optimal medical therapy and left ventricular ejection fraction $<$50\%. Of the 10 HF patients, two also had AF (HF + AF). AF presence and type were based on the most recent device check prior to the date of the CT. All the models used in this paper came from patients who were in sinus rhythm at the time of the CT scan.

\subsection*{Segmentation}

We segmented the LA at the ventricular end-diastolic (ED) CT frame using a semi-automated method within the freely available Seg3D2 software package \cite{CIBC2016Seg3D:Visualization}. The segmentation pipeline implemented here follows that described in \cite{Fastl2018PersonalizedArchitecture}, see Fig \ref{fig:methods_summary}, panel A. Briefly, the ED frame was cropped to include just the LA to reduce the computational overhead and a four-point median filter was applied to the images to enhance image contrast. Here, we identified the LA myocardial tissue, using the assumption that similar Hounsfield unit (HU) intensities exist between atrial and ventricular myocardial tissue \cite{Bishop2016Three-dimensionalFibrillation}. The LA blood pool and myocardium were thresholded within the range of their respective HU means $\pm$ 3 times their standard deviation sampled from 3 manually-selected rectangular regions of interest within the LA bloodpool and LV myocardium, respectively. A sphere was used to crop the LA at the mitral valve (MV) and separate the LA myocardium and blood pool from any additional structures. The pulmonary veins and appendage were cropped from the atrial body perpendicular to the vessel direction\added{.} \deleted{as they were not needed for our simulations, which focused on LA biomechanics only, and their inclusion would increase computational cost.} \added{The contrast-enhanced CT images from which the models were generated fully covered the LA, but in some cases, the appendage was only partially covered and so was excluded from the analysis for consistency. Additionally, this study, and our simulations, focused on LA biomechanics only and aimed to characterise myocardial stiffness of solely the LA body. As such, the pulmonary veins and appendage were not necessary and their inclusion would only increase computational cost.} At the cropped veins, appendage and MV, rings surrounding the edge of the cropped structures were generated. These labels were used as surfaces for boundary conditions in the mechanics simulations. The atrial endocardial surface was then closed using surfaces (here referred to as valve planes) at the MV, pulmonary veins and appendage. This closed endocardial surface was needed to compute the dynamic LA chamber volume during our simulations. The labelled segmentation was smoothed and upsampled to an isometric voxel resolution of 0.15 mm. This reduced the ‘stair-casing’ effect present on the segmentation surfaces and improved the convergence of subsequent mechanics simulations performed with the mesh \cite{Crozier2016Image-BasedModeling}. 

Using Seg3D2, we also segmented the EAT surrounding the LA from the ED CT frame. \deleted{Fig \ref{fig:methods_summary}A (Supporting Information S1 file provides a figure showing the LA EAT for each case)} Here, EAT was automatically identified as a hypo-dense layer with a density between \textminus190 and \textminus30 HU \cite{Monti2020NovelEvaluation}. Since we were only interested in EAT, we then confined the EAT quantified to that within 5 mm of the LA bloodpool. The EAT segmentation was then smoothed and upsampled to an isometric voxel resolution of 0.15 mm, as above.
\added{Fig \ref{fig:methods_summary}A and Supporting Information \nameref{S1_File} shows the LA blood pool, myocardium and EAT segmentation for a representative case. Supporting Information \nameref{S2_File} provides a figure showing the LA EAT for each case.} Some studies have also defined EAT with a range between \textminus195 and \textminus45 HU \cite{ElMahdiui2021PosteriorAblation}, see Supporting Information \nameref{S2_File} for a comparison of the EAT quantification metrics using both HU ranges. 

\subsection*{Mesh generation}

The labelled segmentation was used to generate tetrahedral meshes with the Computational Geometry Algorithm Library \cite{TheCGALProject2023CGALManual}. The atrial anatomies were meshed using tetrahedral elements with maximum edge length of 0.2 mm. This initial edge length was chosen to capture the thin-walled atrial anatomy. The mesh was then resampled using meshtool, an open-source software for mesh processing \cite{Neic2020AutomatingMeshtool}, so that the tetrahedral elements had an average edge length of 0.6 mm. Supporting Information \nameref{S3_File} shows that using the higher resolution mesh of maximum edge length 0.2 mm for simulation resulted in an approximately 28-time fold increase in computation time compared to the lower resolution mesh of average edge length 0.6 mm. The mean difference between the global and regional displacements resulting from simulations at both mesh resolutions was 0.22 mm, which is less than the image resolution of the CT images. As such, all computations in the paper were performed using the lower resolution mesh.  

 The EAT was also meshed with the Computational Geometry Algorithm Library \cite{TheCGALProject2023CGALManual} using tetrahedral elements with maximum edge length of 0.2 mm to capture any thin regions then resampled to an average edge length of 0.6 mm using meshtool \cite{Neic2020AutomatingMeshtool}. The LA chamber and EAT were meshed separately and simulations were carried out using the LA chamber meshes only.

To simulate physiological LA behaviour, we include the influence of the fibre bundles present in atrial anatomy. Using universal atrial coordinates (UACs) defined by Roney et al. \cite{Roney2019UniversalMeshes}, rule-based fibre and sheet orientations were assigned to the model geometries \cite{Solis-Lemus2023EvaluationStudy}.  The fibre and sheet directions used represent histological descriptions of the fibre architecture in a healthy LA \cite{Ho2002AtrialConduction}. These provide an estimation of the preferred myocyte and collagen matrix orientation, that are used when estimating tissue anisotropy. The UACs provide a system by which the complex geometry of the atrial surface is transformed to a 1 unit $\times$ 1 unit square, see Supporting Information \nameref{S4_File}. The endocardial and epicardial surfaces were identified and extracted from the mesh using meshtool and expressed in the UAC system. The rule-based epicardial and endocardial fibres were taken from Labarthe et al. \cite{Labarthe2014AAssessment} and were mapped to their respective surfaces. Transmural fibres were created by projecting the surface fibre architectures through the LA wall using a Laplace based coordinate system, ranging from 0 at the endocardium to 1 at the epicardium \cite{Bayer2018UniversalData}. Epicardial fibres were assigned to the wall regions where the transmural coordinate was greater than 0.5 and endocardial fibres were assigned where the coordinate was less than 0.5. This step-wise transition of the atrial fibres is consistent with histological findings \cite{Ho2002AtrialConduction}. 

Using the UAC system representation of the endocardial surface, the endocardium was split into five regions, namely anterior, posterior, septum, lateral and roof, Fig \ref{fig:methods_summary}B. The vein centres of the right superior and left inferior pulmonary veins in the endocardial UAC system were identified and used to split the the surface into five regions by drawing two perpendicular lines through each vein centre (see Supporting Information \nameref{S4_File}). Each surface mesh element was retagged according to the region in which it belonged. These region element tags were then mapped onto the three-dimensional endocardial surface mesh. Any mislabelled floating elements were corrected using an automated pipeline. Floating elements were located and the region label reassigned to match the region surrounding them. This ensured smooth and continuous region definitions on the three-dimensional surface mesh. The regions defined on the endocardial surface were then projected through the volumetric LA mesh using the transmural Laplace-based coordinate system described previously. 

\subsection*{Quantification of regional left atrial anatomy} 

To estimate the thin and heterogenous thickness of the LA wall, we employed a method based on the eikonal equation and implemented using the cardiac arrythmia research package (CARP) \cite{Vigmond2003ComputationalTissue}. All the vertices on the epicardial surface were activated simultaneously, with the wavefront having a constant, isotropic conduction velocity of 1 mm/s. The thickness of the LA wall was then estimated as the wavefront arrival time at the endocardium. The thickness result was then verified against results from the previously validated but more computationally expensive Laplace-based method described in \cite{Bishop2016Three-dimensionalFibrillation}. The median difference between the average regional thicknesses computed using both methods was 0.13 mm and further detail comparing both methods can be found in Supporting Information \nameref{S5_File}.

\subsection*{In-vivo left atrial deformation} 

The patient-specific model geometries (Figure \ref{fig:methods_summary}, panel A) were generated from the ED frame of 10- or 20-frame ECG-gated CT image sets. Motion models were created using feature tracking to estimate in-vivo LA deformation \cite{Shi2013TemporalDeformations}. These motion models served as the ground truth of the LA deformation in this study. A previously optimised temporal sparse free-form deformation registration method was applied to the gated CT image set to construct a deformation field estimating atrial motion over the cardiac cycle \cite{Sillett2021OptimisationImages}. This was used to transform the ED reference mesh to configurations matching each frame of the gated CT. These 3D motion models were then used to extract several features that described the mechanical behaviour of the LA over the cardiac cycle. This included LA volume and displacement transients. The LA volume transient was used to identify the LA reservoir, conduit and booster pump phases (Fig \ref{fig:methods_summary}B). Global and regional displacement transients of the LA endocardial surface, Fig \ref{fig:methods_summary}B, were also extracted from the image-based motion models. The uncertainty associated with the global and regional  was also calculated from the transformed meshes for each patient. For the volume estimates during the cardiac cycle, we assumed an error of $\pm$5\% patient LA volume at ED.

\subsection*{Biomechanics simulation framework}

All of the mechanical simulations in this study were carried out in CARP \cite{Vigmond2003ComputationalTissue} using 64 cores per simulation in the local high performance computing facilities at King's College London (Tom2) or using 128 cores per simulation on the ARCHER2 UK National Supercomputing Service (https://www.archer2.ac.uk/).

\added{The EAT volume quantified from the CT images was primarily used to investigate correlation with image-derived deformation. EAT was not modelled as a mechanical layer in our simulation framework.}

Orthotropic constitutive laws most accurately reproduce physiological myocardium deformation \cite{Schmid2008MyocardialTests} but use a more complex formulation with a greater number of material parameters than transversely isotropic laws. Thus, it is common to use transversely isotropic material laws with a tractable number of parameters for parameter estimation \cite{Augenstein2005MethodImaging, Wang2009ModellingFunction, Xi2011AnEstimation, Xi2013TheMeasurements, Genet2014DistributionTreatments, Wang2018LeftAnalysis, Babaei2022AEDPVR, Marx2022RobustConfiguration}. In this study, atrial myocardium was modelled as a hyperelastic, nearly incompressible material using the transversely isotropic Guccione law:

\begin{eqnarray}
\label{eq:Guccionelaw}
    \Psi (\textbf{E}) = \frac{C}{2}  \left( e^{Q} - 1 \right) + \frac{\kappa}{2}(\added{\ln}\deleted{\log}(J))^2, 
\end{eqnarray}
where
\begin{eqnarray}
\label{eq:Guccione_Q}
    Q = b_fE_{ff}^2 + 2b_{ft}\left(E_{fs}^2 + E_{fn}^2\right) + b_t\left(E_{ss}^2 + E_{nn}^2 + 2E_{sn}^2\right).
\end{eqnarray}
The $C$ parameter represents the \replaced{isotropic stiffness parameter}{bulk stiffness} of the myocardial tissue while parameters $b_f$, $b_t$ and $b_{ft}$ dictate the mechanical response along the fibre direction, across the transverse planes, and in the fibre-transverse shear planes, respectively. Similar to the reformulation introduced in \cite{Xi2013TheMeasurements}, we included an additional parameter $\alpha$, which scales the anisotropic stiffness parameters: 

\begin{eqnarray}
\label{eq:Guccione_Q_reformulated}
    Q = \alpha \left( b_fE_{ff}^2 + 2b_{ft}\left(E_{fs}^2 + E_{fn}^2\right) + b_t\left(E_{ss}^2 + E_{nn}^2 + 2E_{sn}^2\right) \right).
\end{eqnarray}
The default material parameter values used were $C = 1.7$ kPa, $b_f = $8, $b_{ft}$ = 4, $b_t = $3 as determined in Nasopoulou et al \cite{Nasopoulou2017ImprovedFunction} as estimates for healthy LV myocardium. 
For the non-myocardial tissue, which includes the MV annulus as well as pulmonary vein and appendage rings, the passive mechanics were represented by the non-linear isotropic Neo-Hookean constitutive law:

\begin{eqnarray}
\label{eq:neohookean_law}
    \Psi (\textbf{E}) = c\left(I_1 - 3\right) + \frac{\kappa}{2}\ln^2(J).
\end{eqnarray}
In these tissues, the effect of the fibres is ignored. \added{Physiologically, the pulmonary veins, appendage and MV annulus also exhibit anisotropic behaviour \cite{Holzapfel2000AModels}. However, this is a common simplification that still captures the ability of the vessel to resist deformation under pressure \cite{Holzapfel2000AModels}.} For the pulmonary veins and appendage rings, we used $c=7.45$ kPa  \cite{Strocchi2020ASimulations}. In the valve planes and MV annulus, $c=1000$ kPa to restrict their deformation as the LA moves. In all the tissues, incompressibility was enforced by assigning a bulk modulus of $\kappa=1000$ kPa. 

Omni-directional springs were applied to the right pulmonary veins to constrain LA motion over the simulation. The stiffness of these springs was fixed to 0.001 kPa/$\mu$m. \added{Based on regional atrial motion data obtained from gated CT data from 24 patients \cite{Strocchi2021TheDynamics}, the region of the atrial roof near the right pulmonary veins exhibits limited motion during the cardiac cycle while the region surrounding the left pulmonary veins typically undergoes more deformation. Thus, the left pulmonary veins were not constrained to preserve realistic atrial deformations.} The effect of the pericardium was modelled using normal springs on the surface of the epicardium as described in Strocchi et al. \cite{Strocchi2020SimulatingPericardium}. These normal springs restricted outward displacement only in the direction normal to the epicardial surface while enabling frictionless sliding of the epicardium against the pericardium. \added{The stiffness of the normal pericardial springs varied spatially over the epicardial surface such that the greatest constraint, $k_{peri}$ was on the LA roof and the area of the LA around the MV was free to move. Corresponding to epicardial displacement maps obtained from regional analysis of CT-derived displacements from the atria of 24 patients in \cite{Strocchi2021TheDynamics}, epicardial regions with low and high displacement normal to the surface are applied with the maximum and minimum pericardium penalty, respectively. Across the LA epicardial surface, $k_{peri}$ was scaled according a penalty function, first defined by \cite{Strocchi2020ASimulations} using ventricular motion data. A figure showing an example of the pericardium penalty map of one of the cohort meshes, along with a detailed explanation of the pericardium penalty function can be found in Supporting Information \nameref{S6_File}.}
\deleted{The stiffness of the normal pericardial springs varied spatially over the epicardial surface such that the greatest constraint of 0.001 kPa/$\mu$m was on the LA roof and the area of the LA around the MV was free to move, corresponding to epicardial displacement maps obtained from regional analysis of CT-derived displacements from the atria of 24 patients. A figure showing an example of the pericardium penalty of one of the cohort meshes can be found in Supporting Information S6 File.}

The finite-element method uses a stress-free configuration as the simulation reference state. However, the LA model geometry is constructed from the ED CT image and reflects a pressurized, loaded state. \added{Therefore, to find the unloaded reference stress-free configuration in each simulation, we used an iterative backwards displacement approach as detailed in \cite{Bols2013AVessels}. The inverse problem involves determining the undeformed geometry corresponding to a given geometry that has undergone deformation due to an applied pressure load and boundary conditions. Here, the ED geometry is used as our given reference geometry and an unloaded geometry is calculated for each value of the EDP and stiffness parameters. The forward problem updates the approximate zero-pressure geometry, while evaluating the residual as the difference in chamber volume that is still present between the image-based geometry and the geometry resulting from this forward analysis. This was performed iteratively until a volume-based criteria for convergence was reached, where the approximated unloaded configuration would return a simulated ED configuration within 1.0\% of the image ED configuration volume.  In the unloading simulation, the pericardium and right pulmonary vein constraints are as described previously and the nodes of the MV annulus surface are held fixed by Dirichlet boundary conditions.}

The passive deformation of the LA during its reservoir phase is driven by the force of downward MV motion and increasing endocardial pressure as the chamber fills. Thus, these loading conditions were applied in our simulation model as described below and show in Fig \ref{fig:methods_summary}, panel C. 
 
\added{Due to the lack of patient-specific LA pressure recordings in our cohort, we applied an idealised time-varying pressure boundary condition informed by physiological data reported in the literature \cite{Pagel2003MechanicalEchocardiography}. The simulated pressure profile captures the characteristic phases of passive LA function. A passive loading phase was included at the start of the simulation, during which the LA cavity pressure was increased to end-diastolic pressure (EDP), allowing the chamber to reload to the ED state from its unloaded state. Then during the reservoir phase, pressure was defined to rise from the estimated EDP to end-systolic pressure (ESP), reflecting the passive filling of the LA. This was followed by a decline during the conduit phase. Since atrial contraction was not included in our model, pressure was held constant after the conduit phase. This simplified profile allowed us to replicate the key features of passive LA haemodynamics and is illustrated in Fig \ref{fig:methods_summary}, panel C.}
\deleted{Left atrial pressure recordings were not available for our patient cohort. As such, the time-varying pressure boundary condition used in our simulations followed the shape of a typical pressure profile seen in the LA. During the reservoir phase, LA pressure increases from its ventricular end-diastolic value (EDP) to its peak value at ventricular end-systole (ESP). Following this, in the conduit phase, pressure falls then rises again during atrial contraction. In our simulations, we focused on only passive atrial function, that is reservoir and conduit phases. Therefore, for our simulated pressure profile, we prescribed that the pressure increases from its EDP to its ESP during the filling phase then decreases during the conduit phase then remains constant for the remainder of the cycle, see Fig \ref{fig:methods_summary}, panel C.}

The displacement estimated from the 10 CT frames was interpolated using a third-order spline to obtain a time continuous representation of deformed position of the MV annulus over a cardiac cycle. The interpolated displacement of the MV annulus was applied as time and space varying Dirichlet boundary conditions to each node on the MV annulus mesh surface. Fig \ref{fig:methods_summary}, panel C shows the MV displacement averaged over the nodes on the MV annulus surface for one representative patient. This boundary condition specified the position in space of each node of the MV annulus at each time point in our simulation. The interpolation procedure allowed the simulated displacements to be applied in sufficiently small increments for numerical stability.

\deleted{Therefore, to find the unloaded reference stress-free configuration in each simulation, we used the iterative backwards displacement approach implemented in Marx et al.. In the unloading simulation, the pericardium and right pulmonary vein constraints are as described previously and the nodes of the MV annulus surface are held fixed by Dirichlet boundary conditions. We used a volume-based criteria for convergence where the approximated unloaded configuration would return a simulated ED configuration within 1.0\% of the image ED configuration volume.}

\subsection*{Gaussian process emulation}

Model calibration would require us to evaluate the model at thousands of different parameter values. Using our finite-element simulations for this task would be prohibitively time-consuming and computationally expensive. Instead, we trained Gaussian process emulators (GPEs) to approximate our simulations, aiming to replace the simulations with an approximate but fast to evaluate surrogate model. 

The GPEs used in this study were based on \cite{Longobardi2020PredictingRats} and were implemented using the \verb|GPErks| Python library. The GPEs provide a statistical approximation describing how a scalar simulation output feature varies as a function of the simulation input parameters. Our emulator, $\textbf{f}(\textbf{x})$, was defined as a linear mean function, $\textbf{m}(\textbf{x})$, consisting of explicit basis functions plus a zero-mean Gaussian process, $\textbf{g}(\textbf{x})$:

\begin{eqnarray}
\label{eq:fx}
    \textbf{f} (\textbf{x}) \vcentcolon= \textbf{m} (\textbf{x}) + \textbf{g} (\textbf{x}),
\end{eqnarray}
where $\textbf{x} = (x_1,x_2,\ldots,x_n)^T$ is a vector of input parameters. 
We assume the mean function has the form of a linear regression model

\begin{eqnarray}
\label{eq:mx}
    \textbf{m} (\textbf{x}) = \textbf{h} (\textbf{x})^T \bm{\beta},
\end{eqnarray}
where $\textbf{h}(\textbf{x})$ is the design matrix of predictor variables. We can think of $\textbf{g}(\textbf{x})$ as a non-parametric model that learns the signal accounted for by the mean function, which we model as a zero-mean Gaussian process:

\begin{eqnarray}
\label{eq:gx}
    \textbf{g} (\textbf{x}) \sim \mathcal{GP}(0, k (\textbf{x}, \textbf{x'})).
\end{eqnarray}
We use an exponential quadratic covariance matrix kernel of the form 

\begin{eqnarray}
\label{eq:kxx}
    k (\textbf{x}, \textbf{x'}) = \sigma_f^2 \exp \left( - \sum \left (\frac{\mathbf{x}_i - \mathbf{x}_i'}{\delta_i^2} \right)^2 \right),
\end{eqnarray}
where $\bm{\beta} = (\beta_1,\beta_2,\ldots,\beta_n)$, $\sigma_f$ and $\delta_i$ are hyperparameters to be estimated from the training data.

We trained independent GPEs for each simulation output feature \cite{Longobardi2020PredictingRats}. The number of samples used for GPE training was at least 10 times the input dimension \cite{Loeppky2009ChoosingGuide}.  As such, for the GPE training for each case, we constructed a training dataset comprising simulated outputs for 200 combinations of our 14 input parameters. The points where simulations were performed were selected through Sobol' sampling of our initial parameter space. The input parameters considered and their ranges are shown in Table \ref{tab:sim_inputs}. 

\begin{table}[!ht]

\begin{adjustwidth}{-2.25in}{0in}
\centering
\caption{Simulator input parameters with the input ranges explored during our analysis.}
\begin{tabular}{lcclc}

\hline
Input parameter  &  Units & Range & Definition & References \\
\hline

$C_{anterior}$ & kPa & [0.2, 6.8] & \replaced{isotropic stiffness parameter}{bulk stiffness} in anterior region & \cite{Nasopoulou2017ImprovedFunction} \\

$\alpha_{anterior}$ & - & [0.125, 4.0] &anisotropic scaling in anterior region & \cite{Nasopoulou2017ImprovedFunction}\\

$C_{posterior}$ & kPa & [0.2, 6.8] & \replaced{isotropic stiffness parameter}{bulk stiffness} in posterior region & \cite{Nasopoulou2017ImprovedFunction}\\

$\alpha_{posterior}$   & - & [0.125, 4.0]  &anisotropic scaling in posterior region & \cite{Nasopoulou2017ImprovedFunction} \\

$C_{septum}$ & kPa &[0.2, 6.8] & \replaced{isotropic stiffness parameter}{bulk stiffness} in septum region & \cite{Nasopoulou2017ImprovedFunction}\\
 
$\alpha_{septum}$  & - & [0.125, 4.0] &anisotropic scaling in septal region & \cite{Nasopoulou2017ImprovedFunction}  \\

$C_{lateral}$ & kPa &  [0.2, 6.8] & \replaced{isotropic stiffness parameter}{bulk stiffness} in lateral region & \cite{Nasopoulou2017ImprovedFunction}\\

$\alpha_{lateral}$  & - & [0.125, 4.0] &anisotropic scaling in lateral region & \cite{Nasopoulou2017ImprovedFunction} \\

$C_{roof}$ & kPa & [0.2, 6.8] & \replaced{isotropic stiffness parameter}{bulk stiffness} in roof region & \cite{Nasopoulou2017ImprovedFunction}\\
 
$\alpha_{roof}$  & - & [0.125, 4.0] &anisotropic scaling in roof region & \cite{Nasopoulou2017ImprovedFunction} \\

EDP  & mmHg &  [1, 12]  &LA chamber pressure at ventricular end-diastole & \cite{Braunwald1961LeftCatheterization} \\

ESP   & mmHg & [13, 37] &LA chamber pressure at ventricular end-systole  & \cite{Park2019AtrialPressure}  \\

$k_{peri}$ & kPa/$\mu$m & [0.0001,0.005] &pericardium spring stiffness & \cite{Strocchi2023CellEmulators} \\

PTH  & - & [0.50, 0.95] &pericardium penalty threshold & - \\

\hline
\end{tabular}

\label{tab:sim_inputs}
\end{adjustwidth}
\end{table}

Due to the absence of in-vivo estimates of atrial stiffness properties using the Guccione law, we set the default stiffness values for each LA model using estimated stiffnesses for healthy ventricular tissue from Nasopoulou et al \cite{Nasopoulou2017ImprovedFunction}. The range for these stiffness parameters was set to $\frac{1}{8}$to 4 times the default value to account for the high variability of reported myocardial stiffness values seen across the literature \cite{Xi2011AnEstimation}. The range for the EDP was taken from reported values measured in subjects without cardiovascular disease using a trans-septal catheter \cite{Braunwald1961LeftCatheterization}. For ESP, the range was based on pressure catheter recording taken from AF patients undergoing catheter ablation \cite{Park2019AtrialPressure}. The maximum value of the range of ESP chosen here is higher than peak pressure values reported in advanced heart failure patients \cite{Melenovsky2015LeftFraction}. Thus, we expect that the chosen pressure range should capture appropriate values for the HF patients as well as the those patients with HF and AF. The range of the pericardium stiffness, $k_{peri}$ was chosen to be consistent with previous studies \cite{Strocchi2023CellEmulators}, while the range of the  pericardium penalty threshold (PTH) was varied so that it covered a region starting near the roof of the LA and ending above the MV. For each patient case, the range was adjusted so that the PTH range started at a defined landmark. 

For each simulation, the outputs considered were end-systolic volume (ESV), global and regional displacements ($d_{region}$) where $d_{region}$ is the average displacement of the element centres in the anterior, posterior, septum, lateral and roof region relative to the CT-image ED mesh. The global displacement is taken as the average displacement of the element centres across all of the mesh. This made a total of 7 biomarkers for each simulation. Thus, for each case, 7 GPEs were trained.

GPE predictive accuracy was evaluated using the coefficient of determination ($R^2$) obtained on held-out data using five-fold cross validation. The $R^2$ metric is not robust and can be heavily affected if a small number of points have large errors. We also computed the independent standard error (ISE) as additional metric for GPE accuracy.

$R^2$ serves as an indicator of the accuracy in individual estimates of the GPE posterior mean, with values near 1 indicating minimal discrepancy between predictions and observations:

\begin{eqnarray}
\label{eq:r2}
    R^2 = 1 - \frac{RSS}{TSS}.
\end{eqnarray}
RSS and TSS are the sum of the squared residuals and total sum of squares respectively.
The ISE provides am assessment of the GPE uncertainty. It is defined as the observed average of the 96\% credibility interval, that is, it is the number of data points that lie within 2 standard deviations of the mean prediction:

\begin{eqnarray}
\label{eq:ise}
    \mathrm{ISE}=\frac{\#\left\{i \colon {\mathbb{E}[{{f_i}}\left({\textbf{x}}\right)]}-{{y_i}} <2 \cdot \sqrt{\mathbb{V}ar \left({{f}_i}\left({\textbf{x}}\right)\right)}\right\}}{N},
\end{eqnarray}
where $\textbf{y}$ is the target feature value. 
Thus, an ISE nearing 1 signifies that, for the majority of points, the disparity between the GPE prediction and the corresponding observation falls within the uncertainty range of the GPE. 

After the GPEs were validated, we trained a final GPE for each output using all the simulation data points. The trained GPEs were first used to perform a global sensitivity analysis (GSA) on the our simulation input parameters.

\subsection*{Global sensitivity analysis} 

For the GSA, we carried out a Sobol' variance-based sensitivity analysis \cite{Sobol2001GlobalEstimates} using the Saltelli method  \cite{Saltelli2010VarianceIndex} implemented in the \verb|SALib| Python library \cite{Herman2017SALib:Analysis}. The input parameters were ranked by their importance across all 7 output features. This was done by computing the maximum total effect for each input across all outputs and normalising so that the maximum total effects for each parameter summed to 1, as shown in \cite{Strocchi2023CellEmulators}. 

In our reformulation of the Guccione law (see eq. \ref{eq:Guccione_Q_reformulated}), we found a correlation between $C$ and  $\alpha$ parameters, consistent with previous studies \cite{Xi2013TheMeasurements}. We are not able to uncouple these parameters. Thus, following the GSA, $C$ was fixed at 1.7 kPa for each region and we trained new GPEs corresponding to each of the outputs mentioned above considering only anisotropic stiffness parameters across 5 LA regions ($\alpha_{anterior}$, $\alpha_{posterior}$, $\alpha_{septum}$, $\alpha_{lateral}$, $\alpha_{roof}$ ), LA endocardial pressure at ED and ES (EDP, ESP respectively) and the stiffness of the normal springs for the effect of the pericardium ($k_{peri}$) as well as the threshold over which the PTH was applied.

\subsection*{History Matching}

For each patient, regional  \replaced{isotropic stiffness parameters}{bulk stiffness parameters}, $C_{region}$ were fixed at 1.7 kPa and we calibrated regional $\alpha$ parameters to provide a single index of stiffness per region. 

The history matching (HM) approach, as described in \cite{Coveney2018FittingMatching} and applied to three-dimensional cardiac models in \cite{Longobardi2020PredictingRats, Rodero2023CalibrationMatching, Strocchi2023CellEmulators}, allows us to constrain our initial simulation input parameter space to include only points that reproduce the CT-derived displacement and volume features for each patient. The regional displacements and volume at ES derived from the gated CT image set represent the target values for each patient. HM progresses iteratively in waves with the not-ruled-out-yet (NROY) region of the input space being gradually reduced at each wave. A parameter test set consisting of $100 000$ points was constructed using a Latin hypercube design within the NROY region. The trained GPEs were evaluated at each point within this test set. At each point, $\textbf{x}$, the expected value of the emulator outputs were then compared with the target CT-based observations, $\mathbf{\mu}$, using an implausibility measure, that reflected the discrepancy between emulators’ predictions and data:

\begin{eqnarray}
\label{eq:implausibility}
    I^2 (\textbf{x}) = \underset{i=1,\dots,m}{\max} \frac{(\mathbb{E}[{f}_i (\textbf{x})] - \mu_i)^2}{\V ar [{f}_i (\textbf{x})] + \sigma_i^2}, 
\end{eqnarray}
where $\sigma$ is the standard deviation of the measurement error and $m$ is the number of output features.

Points were then classified as implausible or non-implausible (or NROY) through the introduction of an implausibility threshold, $I_{threshold}$. For a given test input, \textbf{x}, if the maximum implausibility across all output features was greater than $I_{threshold}$, then \textbf{x} was considered implausible. The non-implausible region of the current HM wave then became the NROY parameter space for the subsequent wave. The NROY space was expected to have a complicated geometry after implausible regions were discarded. As such, the Python function \verb|psa_select| from the \verb|diversipy| library was used to generate uniformly distributed samples from this space. The NROY space was sampled so that the simulator was evaluated at $N_{simul}$ points and a new test parameter set of $100 000$ samples was constructed to be used for the next HM wave. For successive waves of HM,  $N_{simul} = 100$.  $I_{threshold}$ was manually initialised to a value of 3.5. $I_{threshold}$ was decremented by 0.5 in each subsequent wave, until it met its final value, $I_{threshold} = 3 $, set according to Pukelsheim's 3 sigma rule \cite{Pukelsheim1994TheRule} then held constant. HM continued until the NROY region converged, that is, until the reduction in the NROY space between successive waves was less than 1\%. 

\subsection*{Markov chain Monte Carlo}

To approximate the posterior distribution of the input parameters and identify an input parameter set most likely to match patient image-based observations, we used Markov chain Monte Carlo (MCMC). MCMC methods provide a way to generate samples from a target probability density. A Markov chain is a sequence of random variables where the probability distribution of each variable depends only on the preceding variable, that is, the future state depends only on the current state, not on the past states. We aim to generate a Markov chain with a stationary distribution of  $\pi(\textbf{x}|\mathbf{y_{\textrm{obs}}})$, representing the posterior distribution of our input parameter $\textbf{x}$ given data $\mathbf{y_{\textrm{obs}}}$.

Our observational model is 
\begin{equation}
    \mathbf{y_{\textrm{obs}}}=\mathbf{F}(\mathbf{x})+\mathbf{\epsilon}
\end{equation}
where $\mathbf{y_{\textrm{obs}}}$ is our observation (analogous to $\mathbf{\mu}$ in the HM context), $\mathbf{F}$ is our observational model, $\mathbf{x}$ are the model parameters and $\epsilon_i$ is a normally distributed observation error, $\epsilon_i\sim N(0,\sigma_i^2)$. For a given observation, $\mathbf{y}_{\textrm{obs}}^*$, we seek the posterior distribution $\pi(\textbf{x}|\mathbf{y}_{\textrm{obs}}^*)$. Under a Bayesian framework this is proportional to $\pi(\textbf{x})\pi(\mathbf{y}_{\textrm{obs}}^*|\textbf{x})$ where $\pi(\textbf{x})$ is the prior distribution over $\mathbf{x}$ and $\pi(\mathbf{y}_{\textrm{obs}}^*|\textbf{x})$ is the likelihood. As $\mathbf{F}(\textbf{x})|\textbf{x}$ and $\mathbf{\epsilon}$ are normally distributed, the likelihood is given by

\begin{equation}
    \pi(\mathbf{y}_{\textrm{obs}}^*|\textbf{x}) = \prod_{i=1}^m\frac{1}{\sqrt{2\pi\left(\sigma_i^2+ \V ar [{f}_i (\textbf{x})]\right)}}\exp{\left(- \frac{1}{2(\sigma_i^2+\V ar [{f}_i (\textbf{x})])}(\mu_i -\mathbb{E}[{f}_i (\textbf{x})])^2\right)},
\end{equation}
using the same notation as in the GPE section. This likelihood accounts for both the observation error, $\sigma_i$, and the variance of the GPE, $\V ar [{f}_i (\textbf{x})])$.

For our analysis, we used the \verb|emcee| python library \cite{Foreman-Mackey2013EmceeHammer}, designed to run efficiently with minimal user interaction. The \verb|emcee| package is a Python implementation of Goodman and Weare's affine Invariant MCMC Ensemble sampler \cite{Goodman2010EnsembleInvariance}. This method employs a number of Markov chain walkers that all know where each other is and update accordingly. 

Using the \verb|emcee| library, we implemented an ensemble sampler of 18 parallel walkers of $100 000$ steps. We used uniform prior distributions for each of the 9 input parameters, defined within the ranges derived following the final wave of HM. A burn-in period of $10 000$ steps was used to discard the initial steps dependent on the starting position of the Markov chain. We used a thinning frequency of 10 to reduce the effect of correlated samples \cite{Riabiz2022OptimalOutput}. In cases where the MCMC chain did not converge after $100 000$ steps, we increased the number of steps and the length of the burn-in period accordingly. 

\subsection*{Linear Mixed Effect Models}

\added{Our data contains repeated measures from each patient case (one sample per region) and therefore are not independent. Thus, we used mixed effect models to examine the relationship between regional LA displacement and LA properties (regional wall thickness, EAT volume and tissue stiffness) while taking into account that our data consists of repeated samples within a single patient case as well as samples across patients. Linear mixed effect models extend traditional linear models to include both random and fixed effects as predictor variables. This allows for non-independence, such as that which arises with repeated measurements, within the data to be explicitly modelled \cite{Harrison2018AEcology}. Fixed effects relate to relationships at the population level while random effects relate to a grouping variable and allow estimation of how the response variable varies within and across groups \cite{Breslow1993ApproximateModels}.}

\added{To set up our mixed effect models, we used the \texttt{statsmodels} python library \cite{Perktold2024Statsmodels/statsmodels:0.14.2}. To determine if a predictor variable significantly impacted the response, we used a significance level of 0.05. For p-values less than or equal to 0.05, we could conclude that a given factor was a significant determinant of LA biomechanics. P-values were derived using a Wald test that involves comparing the estimated value of the coefficient, $\beta_i$, with the estimated standard error for the coefficient, as implemented in the \texttt{statsmodels} python library \cite{Perktold2024Statsmodels/statsmodels:0.14.2}.}

\deleted{To elucidate the relationship between LA biomechanical function, encoded in regional LA displacement, and the LA passive properties, including regional wall thickness, EAT volume and tissue stiffness, we used linear mixed effect models. Linear mixed effect models extend traditional linear models to include both random and fixed effects as predictor variables and allow any correlation in the data to be explicitly modelled. Random effects relate to some grouping variable and allow estimation of how the response variable varies within and across groups. Our data contains repeated measures from each patient case (one sample per region) and therefore are not independent. Thus, we used mixed effect models to examine the relationship between regional LA displacement and LA properties (regional wall thickness, EAT volume and tissue stiffness) while taking into account that our data consists of within patient samples as well as samples across patients.}

The linear mixed-effects model was formulated as:

\[
Y_{ij} = \beta_0 + \beta_1 \cdot \text{C(region)}_j + \beta_2 \cdot X_{ij} + u_j + \varepsilon_{ij}
\]
where \(Y_{ij}\) is the response variable for observation \(i\) within group \(j\), \(\beta_0\) represents the fixed intercept, \(\beta_1 \cdot \text{C(region)}_j\) is the fixed effect for the categorical predictor $\textrm{region}$, \(\beta_2 \cdot X_{ij}\) is the fixed effect for the continuous predictor corresponding to the LA property of interest (regional wall thickness, EAT volume or tissue stiffness), \(u_j\) is the random intercept for each group \(j\) and \(\varepsilon_{ij}\) is the residual error, assumed to be independently and identically distributed as \(\mathcal{N}(0, \sigma^2)\).  Please note that the notation in this section is independent of previous sections. In this model, the observations was grouped by patient case. 

\deleted{To set up our mixed effect models, we used the \texttt{statsmodels} python library. To determine if a predictor variable significantly impacted the response, we used a significance level of 0.05. For p-values less than or equal to 0.05, we could conclude that a given factor was a significant determinant of LA biomechanics.}

\bigskip
The code to perform the analysis from meshing to MCMC analysis is available on github: 
\url{https://github.com/CEMRG-publications}.

\section*{Results}

In this paper, we aimed to investigate the effect of anatomical features and myocardial material properties on atrial deformation. First, we present results from a gated CT image based analysis where we looked at the contribution of image derived regional anatomical features to the LA mechanical behaviour observed from the images. Next, we present results of our stiffness calibration and validation approach performed using passive mechanical simulations and GPEs. Finally, we examine the relationship between regional stiffness parameters, anatomy and LA biomechanics.

\subsection*{Cohort properties}

\begin{figure}[H]
    \centering
    \includegraphics[width=\linewidth]{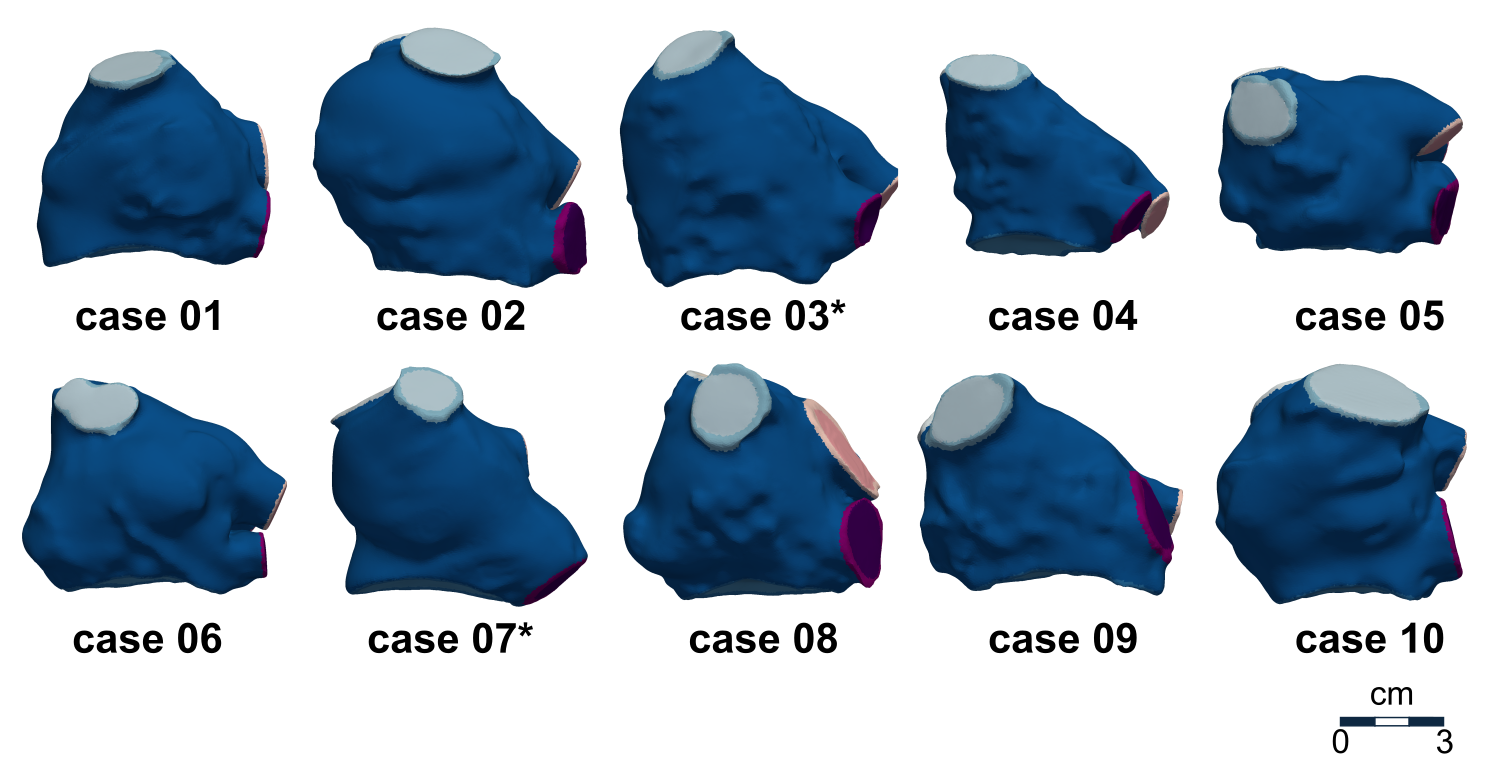}
    \caption{\textbf{Left atrial mesh cohort}. Images show the anterior view of the cohort of 10 meshes generated using the pipeline described in Methods. * indicates patients with AF. }
    \label{fig:cohort}
\end{figure}

Our cohort of patients included 9 males and 1 female with average age 67 ± 13 years.
Table \ref{tab:demographics} shows demographics of the patient cohort and Fig \ref{fig:cohort} shows the cohort of patient-specific LA models used in this study.

\begin{table}[!ht]

\begin{adjustwidth}{-2.25in}{0in}
\centering
\caption{Patient cohort demographics.}
\begin{tabular}{lccccccc}
\hline
\multicolumn{1}{c}{} & \begin{tabular}[c]{@{}c@{}}Age\\   {[}years{]}\end{tabular} & \begin{tabular}[c]{@{}c@{}}Sex\\  {[}-{]}\end{tabular} & \begin{tabular}[c]{@{}c@{}}AF\\   {[}-{]}\end{tabular} & \begin{tabular}[c]{@{}c@{}}ED\\  volume\\  {[}ml{]}\end{tabular} & \begin{tabular}[c]{@{}c@{}}No. of \\  CT frames\\   {[}frames{]}\end{tabular} & \begin{tabular}[c]{@{}c@{}}Heart rate \\  during scan\\   {[}bpm{]}\end{tabular} & \begin{tabular}[c]{@{}c@{}}CT resolution\\  {[}mm x mm x mm{]}\end{tabular} \\
\hline
case 01 & 82 & M & \texttimes & 62 & 20 & 60 & 0.32 x 0.32 x 0.50 \\
case 02 & 63 & M & \texttimes & 117 & 20 & 40 & 0.41 x 0.41 x 0.50 \\
case 03 & 72 & M & $\checkmark$ & 135 & 10 & 70 & 0.45 x 0.45 x 0.50 \\
case 04 & 49 & M &\texttimes & 69 & 20 & 100 & 0.39 x 0.39 x 0.50 \\
case 05 & 76 & M &\texttimes & 72 & 10 & 51 & 0.42 x 0.42 x 0.40 \\
case 06 & 67 & M & \texttimes & 102 & 10 & 70 & 0.49 x 0.49 x 0.40 \\
case 07 & 83 & M & $\checkmark$ & 96 & 10 & 100 & 0.44 x 0.44 x 1.0 \\
case 08 & 69 & F & \texttimes & 113 & 10 & 60 & 0.41 x 0.41 x 0.40 \\
case 09 & 41 & M & \texttimes & 47 & 10 & 50 & 0.41 x 0.41 x 0.40 \\
case 10 & 68 & M & \texttimes & 82 & 20 & 65 & 0.37 x 0.37 x 0.50 \\

\hline
\end{tabular}

\label{tab:demographics}
\end{adjustwidth}
\end{table}

\subsection*{Image-derived LA motion}

\begin{figure}[H]
    \centering
    \includegraphics[width=10cm]{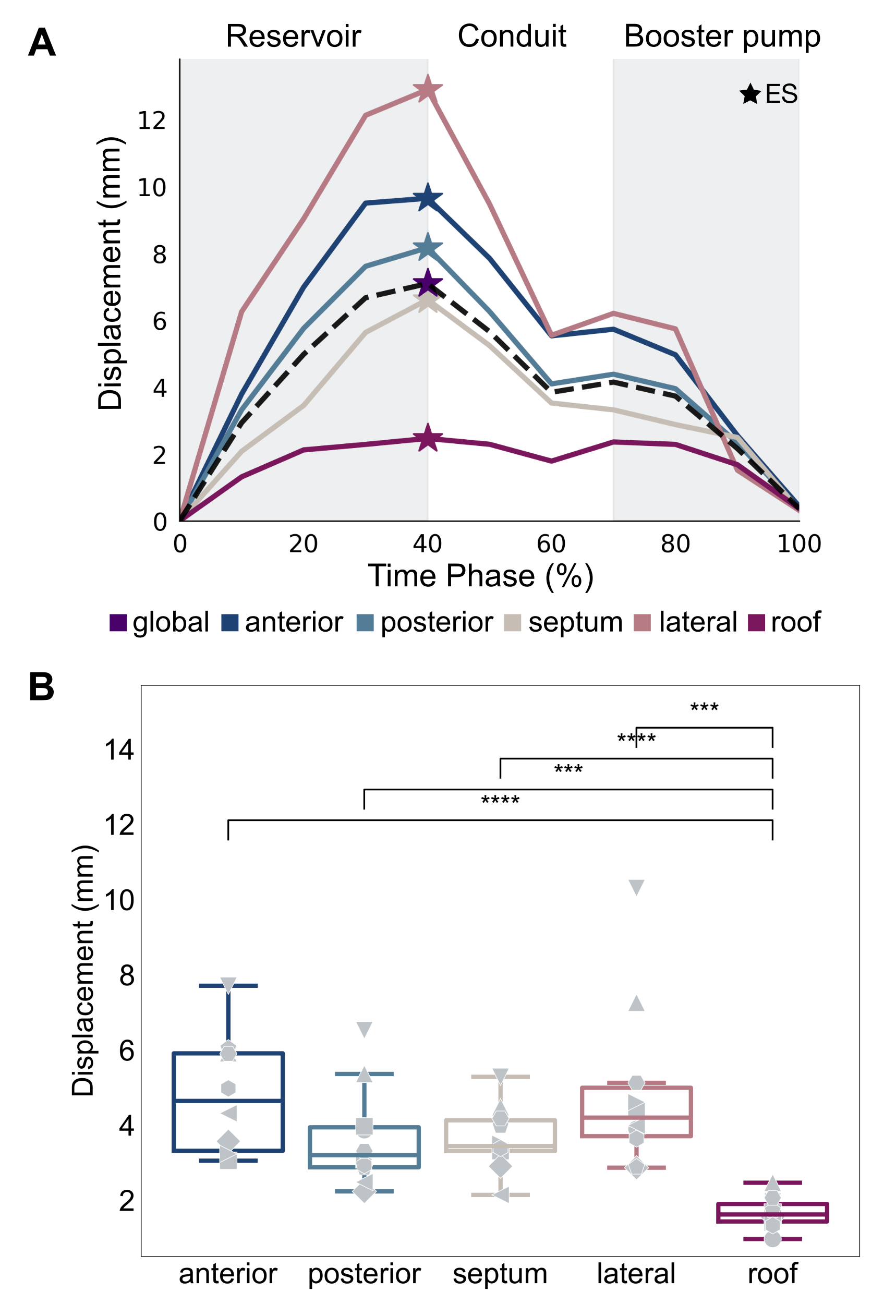}
    \caption{\textbf{Regional heterogeneity in LA displacement}. Plots show regional endocardial surface displacements transients derived from the feature tracking motion models for a representative case (A) and the distribution of the regional ES displacements across the cohort (B). Each marker symbol corresponds to one of the 10 patients. Significance was determined using a paired t-test with the Bonferroni correction for multiple comparisons applied. *** indicates a p-value $<$ 0.001. **** indicates a p-value $<$ 0.0001. }
    \label{fig:image_motion}
\end{figure}

Through feature tracking, we developed motion models of the LA, such that one mesh was produced at each frame of the gated CT image-set. This provided patient-specific estimates of physiological LA deformation. From this, we saw that LA displacement at ES varies across LA regions  (Fig. \ref{fig:image_motion}, panel A) with the roof deforming significantly less than other regions ($p < 0.001$) (Fig. \ref{fig:image_motion}, panel B).

\subsection*{Correlation of regional left atrial anatomy with in-vivo deformation}

The complex, heterogeneous anatomy of the LA as well as the surrounding volume of EAT might affect LA mechanical behaviour and deformation. To investigate the impact of LA geometry on its mechanical function, we examined the correlation between regional thickness and regional EAT volume at the reference configuration with regional displacement derived from the CT image set. We used mixed effect models to examine the relationship between displacement and anatomy.

\begin{figure}[H]
    \centering
    \includegraphics[width=\linewidth]{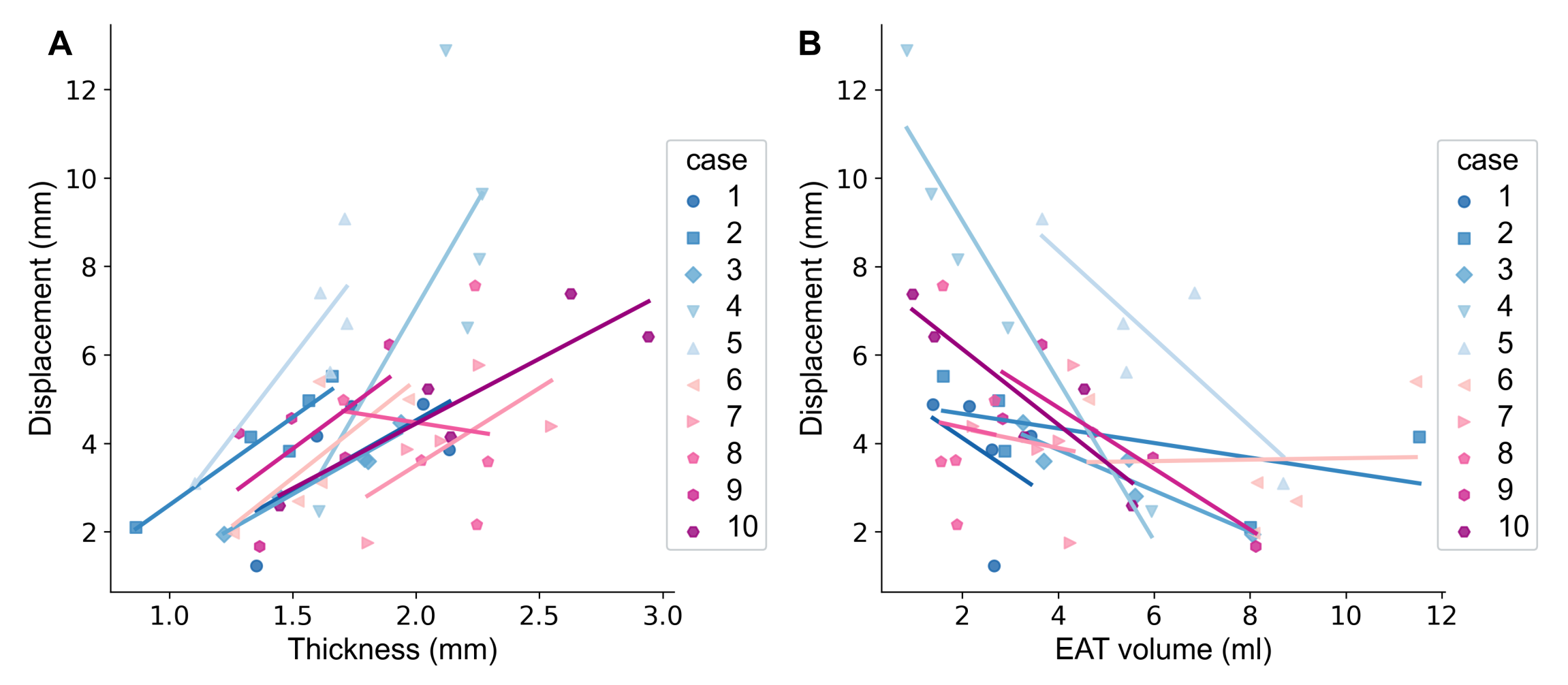}
    \caption{\textbf{Effect of anatomical features on observed LA displacement}. Plots show how regional averages of end-systolic (ES) displacement vary with regionally averaged LA wall thickness (A) and regional EAT volume (B). Each marker symbol corresponds to one of the 10 patients. }
    \label{fig:anatomy_correlations}
\end{figure}

Fig \ref{fig:anatomy_correlations} shows plots of regional displacement at ES against regional wall thickness (panel A) and EAT (panel B) for our 10-patient cohort.

In our mixed effect model, we considered LA region and regional wall thickness as independent factors dictating regional displacement. This model showed that ES displacement changes with region, however, LA wall thickness does not have a significant effect on LA displacement at ES ($p = 0.365$). Using a similar approach to look at the relationship between EAT and ES displacement, we saw that across the LA, regional EAT volume did not have a significant effect on regional displacement at ES ($p = 0.228$).

Based on these results, LA anatomy and surrounding EAT do not have a significant role in determining LA deformation. We hypothesise that myocardial stiffness and/or boundary conditions might contribute to heterogeneous chamber biomechanics over factors such as EAT volume and wall thickness.

\subsection*{Passive stiffness sensitivity analysis}

\begin{figure}[H]
    \centering
    \includegraphics[width=\linewidth]{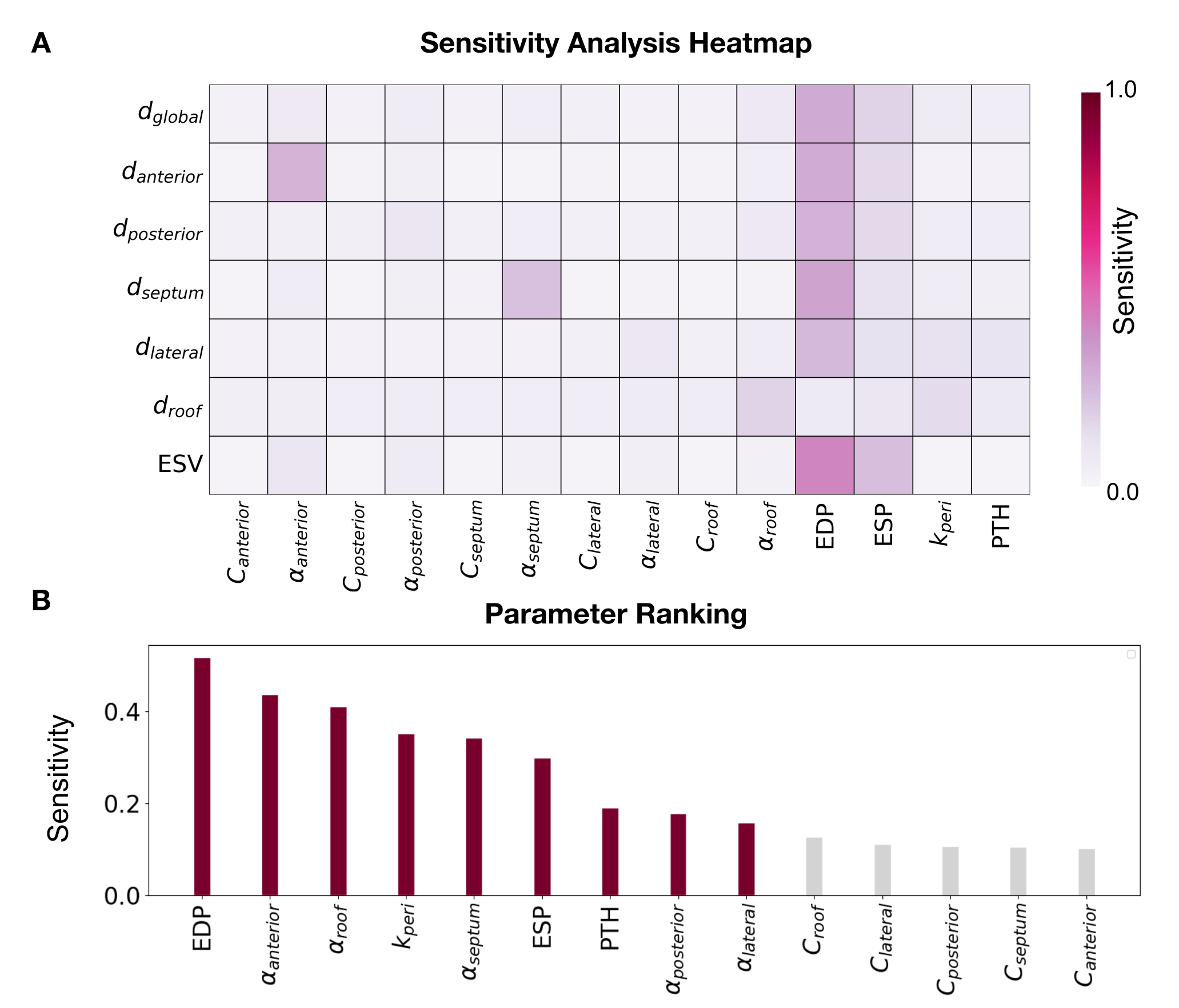}
    \caption{\textbf{Global sensitivity analysis results}. \textbf{A} Heatmap of the total effect of the parameters (x-axis) on the outputs (y-axis). \textbf{B} Barplot of the maximum total effect of each parameter over all outputs. The parameters are ranked from most to least important. The coloured bars represent the total sensitivity up to 90\%.}
    \label{fig:gsa}
\end{figure}

Fig \ref{fig:gsa}A shows the heatmap for the total effect of all input parameters (columns) over all outputs (rows) for one representative case, with dark colors showing high interaction between inputs and outputs.  Fig \ref{fig:gsa}B presents a barplot ranking the input parameters from most to least important based on the maximum total effect of each parameter over all outputs for one representative patient case. See Supporting Information \nameref{S7_File} for individual patient heatmaps and parameter rankings for the whole cohort. In each region, $\alpha_{region}$ outranks the $C_{region}$ parameter, suggesting that the $\alpha_{region}$ parameters are more relevant to regional deformation than regional $C$ parameters.

Based on this GSA, for the regional stiffness calibration, we focused on fitting $\alpha$ values for each of the 5 regions. The results presented in the following sections pertain to the simulation framework where regional $\alpha$ parameters, EDP, ESP, pericardial stiffness and PTH are considered as input parameters; $C_{region}$ was held fixed at 1.7 kPa, the reference value obtained from \cite{Nasopoulou2017ImprovedFunction}.

\subsection*{Calibration of model parameters}

\begin{figure}[H]
    \centering
    \includegraphics[width=\linewidth]{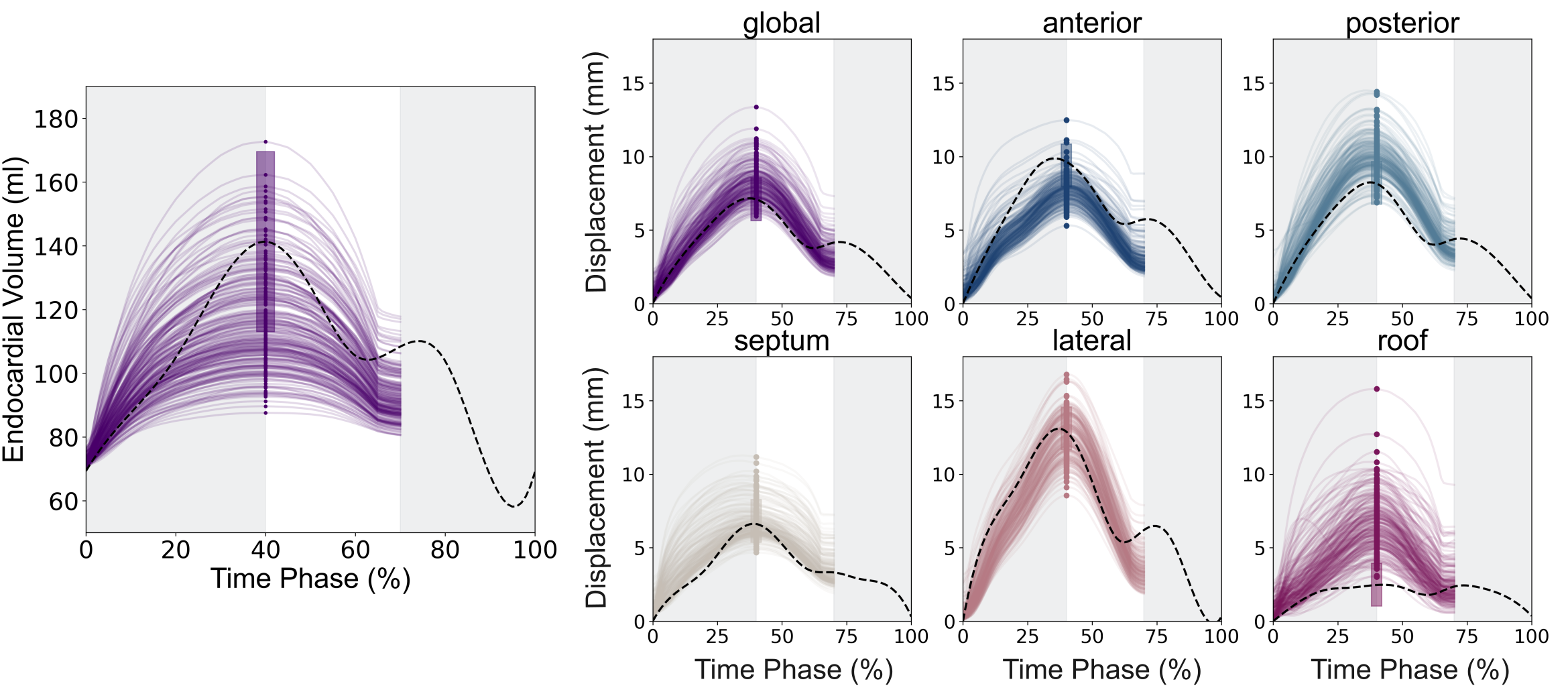}
    \caption{\textbf{Simulated output transients}. The panels show the global and regional displacements as well as the volumes obtained from the initial simulator dataset. The dashed black line indicates the LA behaviour derived from the CT image set. The solid circles indicate the time of ventricular end-systole (ES). The LA reservoir, conduit and booster pump phases were identified from the volume transient. Our simulations focused on passive LA function, hence only the reservoir and conduit phase were simulated. }
    \label{fig:wave1_outputs}
\end{figure}

For each case, 200 successfully completed simulations formed the training dataset for the GPEs with one GPE being trained for each output feature. Fig \ref{fig:wave1_outputs} shows the simulated regional displacement curves and volume transients achieved by the left atrial simulation framework for one representative case. The ventricular ES time point was determined to occur at the peak of the volume transient and is indicated by the coloured dots in Fig \ref{fig:wave1_outputs}. The CT-derived regional displacement and volume curves for this patient are represented by the dotted lines in each panel. Our simulation framework was plausibly able the reproduce the patient's LA volume and regional displacements at end-systole. 

 We recorded the mean $R^2$ and ISE scores for each output. Across the cohort, over all the outputs, the $R^2$ and ISE scores were all greater than 0.72 and 0.89, respectively (full details on the results of emulator training can be found in Supporting Information \nameref{S8_File}). This suggests that the emulators were able to give statistically reasonable estimates of all the desired outputs across our patient cohort. 

\begin{figure}[H]
    \centering
    \includegraphics[width=\linewidth]{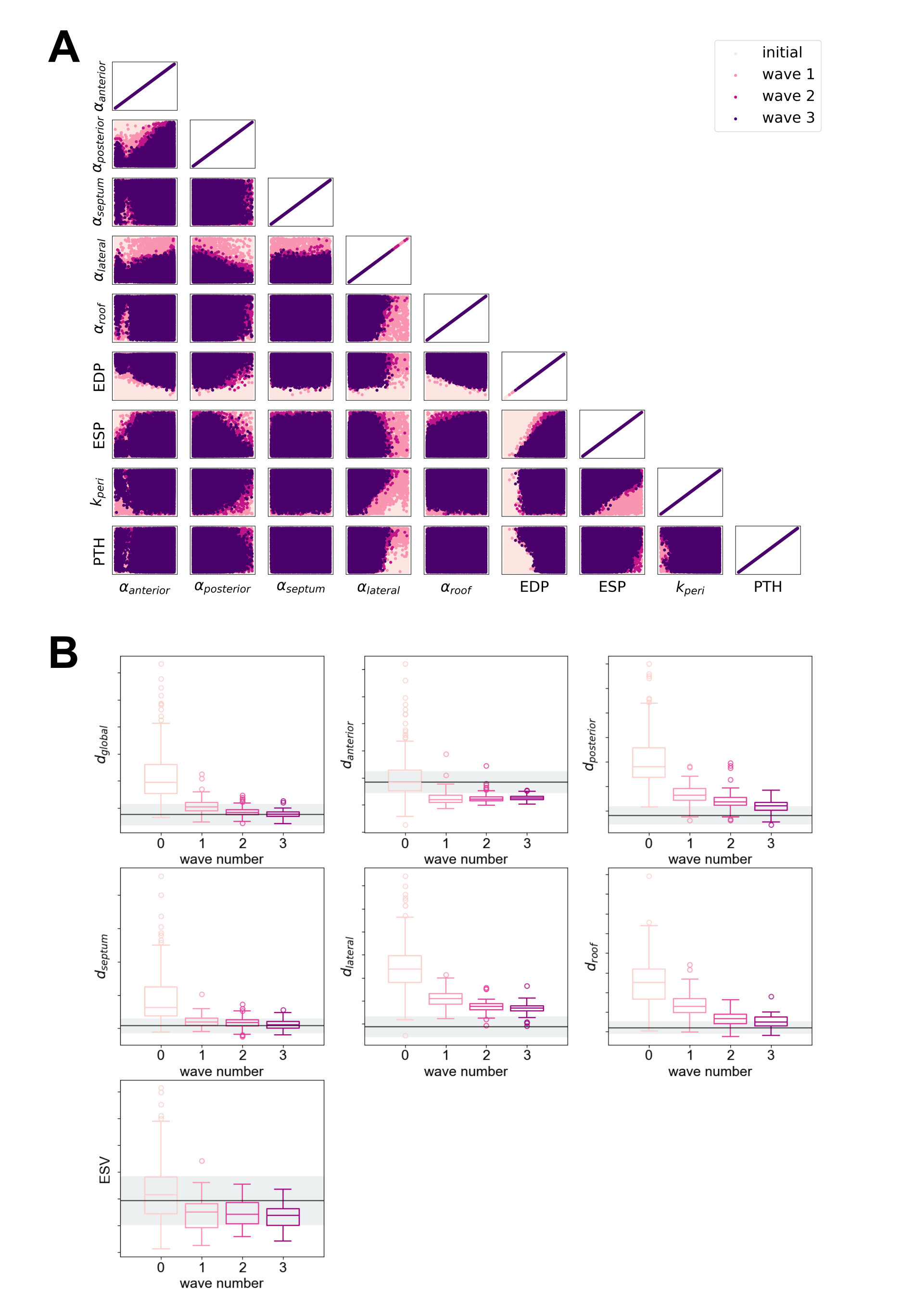}
    \caption{\textbf{History Matching}. (A) High-dimensional input parameter space reduction during HM. (B) Changes in the simulator outputs for each wave of history matching. The solid horizontal lines indicate the value of the output feature at ventricular end-systole (ES) estimated from the CT images. The shaded areas a represent 95\% confidence interval on the observed output feature values. For $d_{region}$, the standard deviation is estimated from feature tracking and for ESV, the standard deviation is assumed to be $\pm$5\% EDV. }
    \label{fig:HM}
\end{figure}

HM was applied in each patient case with inital parameter ranges taken from Table \ref{tab:sim_inputs}. Fig \ref{fig:HM}, panel A plots the nine-dimensional input space as a two-dimensional projection for each pair of parameters and illustrates the multi-dimensional space reduction achieved through iterative HM waves. The initial input set, shown as the lightest colour in Fig \ref{fig:HM}, consisted of 100 000 points sampled using Latin hypercube sampling over the input space. HM constrained the input parameter space to a reduced but still broad range of the parameter space, shown as the darkest colour in Fig \ref{fig:HM}. Across the cohort, the percentage reduction in the initial input space ranged from 45\% to over 99\%. Supporting Information \nameref{S9_File} provides further detail on the NROY space reduction at each wave of HM for each case.

As HM reduced the input space, the simulator outputs obtained using inputs sampled from within the NROY converged towards the values observed from the patient clinical data for all the output features considered as shown in Fig \ref{fig:HM}, panel B.

Through HM, we identified a reduced region of the input space that allowed for accelerated parameter calibration using MCMC.

\subsection*{Calibration of model parameters using MCMC}

\begin{figure}[H]
    \centering
    \includegraphics[width=\linewidth]{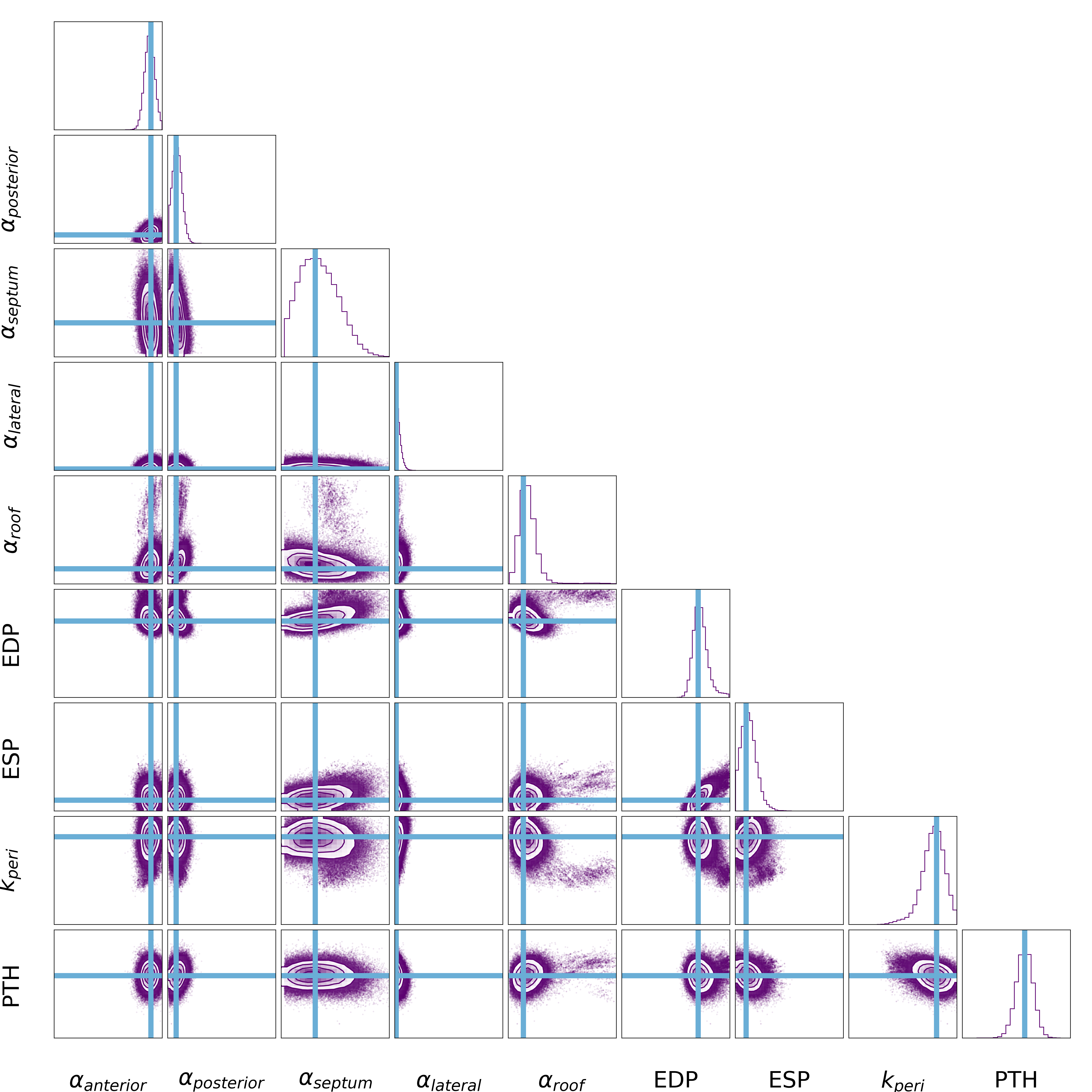}
    \caption{\textbf{Calibration using MCMC}. Plots showing the parameter distributions for a representative patient estimated using MCMC. Each tile represents a projection of the Markov chain samples. We implemented an ensemble sampler of 18 parallel walkers of 100 000 steps. A burn-in period of 10 000 steps and a thinning frequency of 10 was used. For each parameter, the MAP estimate is indicated by the blue line. The histograms along the diagonal display the posterior distributions for each parameter individually. The off-diagonal panels show the 2D distributions as contour plots where the contours are drawn at levels containing 11.8\%, 39.3\%, 67.5\% and 86.4\% of the samples. }
    \label{fig:cornerplot}
\end{figure}

The posterior distribution of the input parameters, estimated using MCMC, is shown for a representative case in Fig \ref{fig:cornerplot}. The maximum a posterior (MAP) estimate obtained from the MCMC sampler represents the set of input parameters with the highest posterior probability of returning the target CT-derived features for each case. For each parameter, the MAP is indicated by the blue line in Fig \ref{fig:cornerplot}. See Supporting Information \nameref{S10_File} for the plots showing the posterior distribution for each patient case.

\begin{figure}[H]
    \centering
    \includegraphics[width=\linewidth]{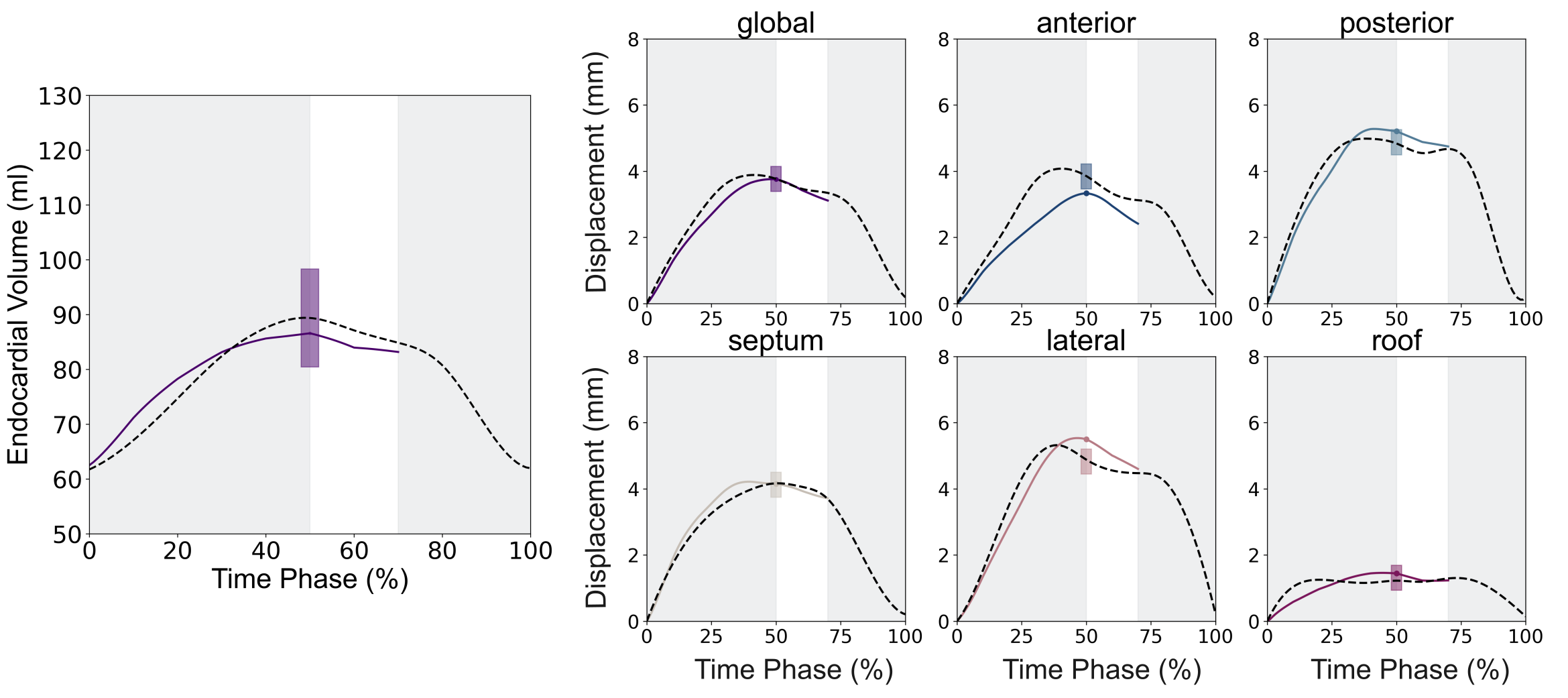}
    \caption{\textbf{Calibrated outputs}. The panels show the global and regional displacements as well as the volumes obtained when the simulator was evaluated using the MAP dataset. The dashed black line indicates the LA behaviour derived from the CT image set and the shaded box represents 95\% confidence interval on the target value. }
    \label{fig:calibrated_outputs}
\end{figure}

    Using the samples obtained from the final wave of HM, we found the nearest neighbour to the MAP estimate obtained using MCMC for each case. We saw good agreement between the simulated and observed output features as shown in Fig \ref{fig:calibrated_outputs}. Over our patient cohort, the average root mean squared error (RMSE) between the simulated and patient-derived ESV was 11.0 ml, approximately 9 $\%$ of the average LA volume of the cohort. The average RMSE between the simulated and target displacements was 0.49 mm globally and 1.55 mm, 1.16 mm, 0.86 mm, 0.73 mm, and 0.62 mm in the anterior, posterior, septum, lateral and roof regions respectively. Following calibration, the simulated global and regional displacement transients were validated against the image-derived displacements through the reservoir and conduit phase, as shown in Supporting Information \nameref{S11_File}.

\begin{figure}[H]
    \centering
    \includegraphics[width=\linewidth]{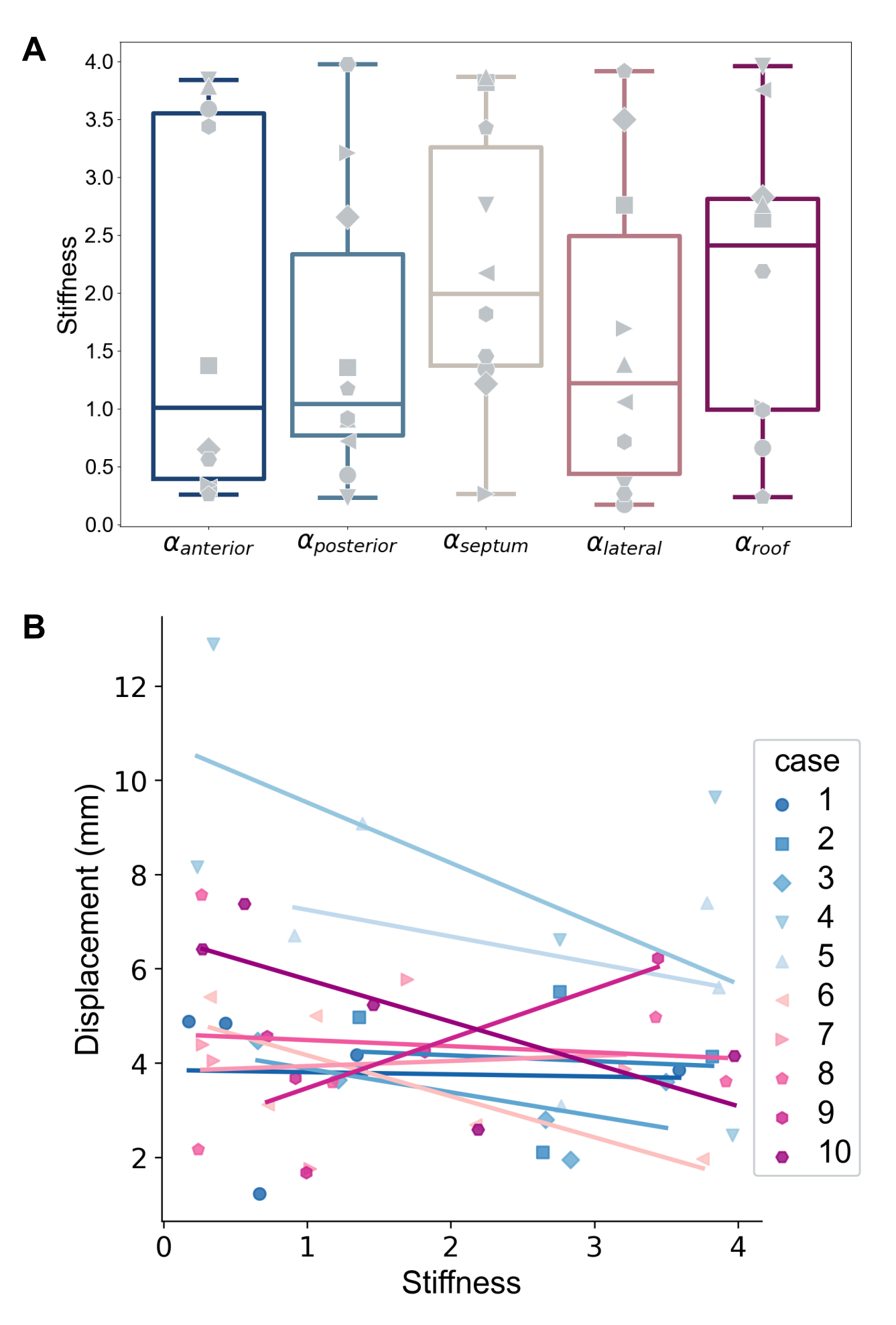}
    \caption{\textbf{Regionally calibrated stiffness parameters}. (A) Box plots showing the distribution of calibrated stiffness parameters per region across the 10-patient cohort. Each marker symbol corresponds to one of the 10 patients. (B) Plots show how regional averages of end-systolic (ES) displacement vary with regional stiffness. }
    \label{fig:stiffness_correlations}
\end{figure}

With the calibrated regional stiffness parameters for each case, we investigated if any trends in regional stiffness existed in the cohort. Fig \ref{fig:stiffness_correlations}, panel A suggests that regional stiffness heterogeneity exists in each patient case. However, across the patient cohort, there was no region where the stiffness was consistently higher relative to the other regions. Using a paired t-test, with a Bonferroni correction for multiple comparisons, we found no consistent and systematic difference between regional stiffness across the cohort.

 Using mixed effects models, we evaluated the relationship between region category, stiffness and observed ES displacement. In our mixed effect model, we assumed that LA region and regional stiffness were independent factors dictating regional displacement. This implies that the effect of stiffness on regional ES displacement does not differ significantly across the different regions. We found that myocardial stiffness plays a significant role in determining LA displacement at ES ($p=0.023$). We also used linear mixed effect models to investigate the relationship between the regionally calibrated stiffness parameters and the anatomical features of interest. We found no significant association between stiffness and regional wall thickness ($p=0.127$) or regional EAT volume ($p=0.205$).

\subsection*{Verification of biomechanical simulation, emulation and calibration framework}

 \added{The contraction of the posterior distribution towards the image-derived output features shown in Supporting Information \nameref{S10_File} provides some indication that HM and MCMC calibration pipeline is working as intended. However, in this section, we perform a systematic verification of our modelling framework using synthetic data. }

 \added{Using case 01, we selected an input parameter set near the centre of the input space, with ranges as defined previously (see Table \ref{tab:noise_error}). For this synthetic data, $C_{region} = 1.7$ kPa, as set following the GSA. We used this selected parameter set as inputs for our full finite-element biomechanics simulation. The outputs of this simulation then became our synthetic target output data. This synthetic data comprised the same 7 features (ESV, $d_{anterior}$,  $d_{posterior}$,  $d_{septum}$, $d_{lateral}$, $d_{roof}$) that we obtained from the CT image data that we previously tried to match.}

\added{To calibrate model input parameters, we followed the HM and MCMC approach outlined above. With each successive HM wave, we attempted to match the synthetic data as opposed to the CT-derived data. The uncertainty attached to our image-derived displacement targets was estimated from discrepancies in the feature tracking when returning the ED LA configuration. For this synthetic data, we set the expected standard deviations on the displacement targets to be 0.2 mm. This aligned with the image-derived uncertainty for this case. For ESV, the standard deviation was set to 5\% of the target value, following that set in the main analysis. Following 3 waves of HM, the initial input space was reduced by 70\% but initial parameter ranges were not reduced.}
 
\begin{figure}[H]
    \centering
    \includegraphics[width=\linewidth]{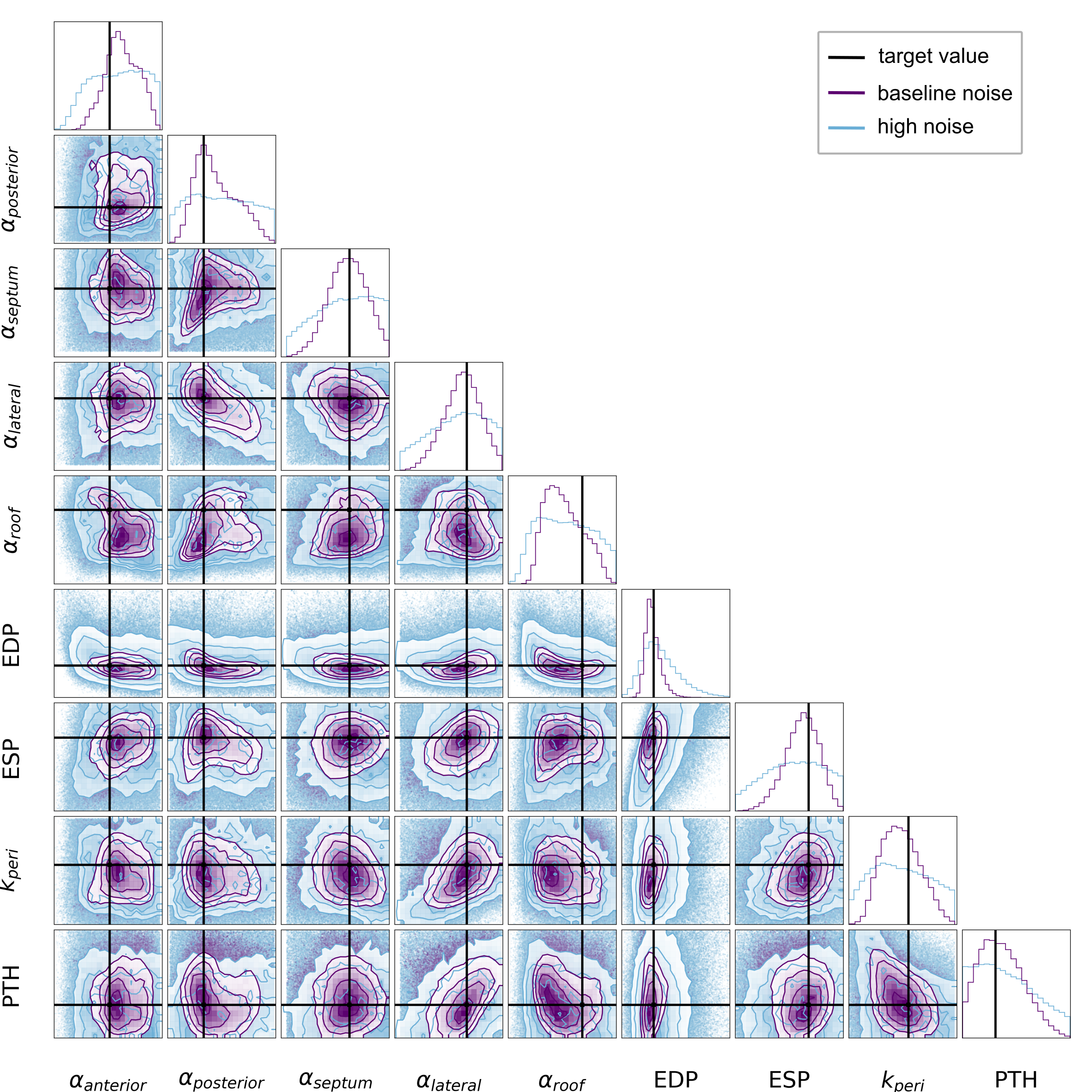}
    \caption{\textbf{MCMC Calibration using synthetic data for case 01}. The MAP estimates returned by MCMC at the lowest and higest level of noise are indicated by the purple and blues lines respectively. The target input values are shown in black. The histograms along the diagonal display the posterior distributions for each parameter individually. The off-diagonal panels show the 2D distributions as contour plots where the contours are drawn at levels containing 11.8\%, 39.3\%, 67.5\% and 86.4\% of the samples. }
    \label{fig:mcmc_verification}
\end{figure}

\added{We then performed the MCMC analysis using the \texttt{emcee} library and implemented an ensemble sampler of 18 parallel walkers of 100 000 steps with a burn-in period of 10 000 steps and a thinning frequency of 10. The cornerplot shown (Figure \ref{fig:mcmc_verification}) illustrates the performance of the MCMC method in recovering the known input parameter values for case 01 (purple plots).}

\added{To ensure that the fitting procedure was robust to increasing amounts of noise, likely to be present in medical data, we analysed the effect of increasing uncertainties attached to the target data on the calibrated input values. The expected standard deviations on the displacements were increased from 0.2 mm to 1.0 mm, which represented the maximum discrepancy coming from feature tracking in our cohort. Uncertainties on ESV were increased from 5\% to 20\%. The effect of increased noise on MCMC calibration can be seen in Fig \ref{fig:mcmc_verification} (blue plots).}

  \added{Generally, the MAP estimates did well at recovering the known target values across all the input parameters, Fig  \ref{fig:mcmc_verification}. There was a larger discrepancy between the MAP estimate and the true value of $\alpha_{roof}$ as can be observed in Fig \ref{fig:mcmc_verification}, however the target value still fell within the posterior distribution. The observation error, or noise, attached to the target features did not appear to create substantial variation in the predicted input parameter values. Table \ref{tab:noise_error} shows that the distance of the MAP from the target value was largely maintained in the presence of increasing noise. The posterior distributions plotted in Fig \ref{fig:mcmc_verification} provide a measure of the uncertainty attached to the MCMC predictions. With increasing noise, the uncertainty of the predicted values increased (wider distributions) but the MAP values remained relatively consistent (Table \ref{tab:noise_error}).}

\begin{table}[!ht]
\begin{adjustwidth}{-2.25in}{0in}
\centering
\caption{\textbf{Summary of verification study results for input parameters}. The table lists each simulator input parameter with its target value and the corresponding uncertainty and accuracy metrics under two observation noise scenarios: baseline noise and high noise. For each noise condition, we report the 95\% confidence interval (CI) of the posterior distribution, whether the target value lies within this interval, and the distance of the MAP estimate from the target.}
\label{tab:noise_error}

\footnotesize
{\renewcommand{\arraystretch}{1.2}
\begin{tabular}{lcccccccc}
\hline
\rule{0pt}{3.2ex} 
\multirow{3}{*}{\begin{tabular}[c]{@{}c@{}}Input \\ parameter\end{tabular}} &
\multirow{3}{*}{Units} &
\multirow{3}{*}{\begin{tabular}[c]{@{}c@{}}Target \\ value\end{tabular}} &
\multicolumn{3}{c}{\textbf{Baseline noise}} &
\multicolumn{3}{c}{\textbf{High noise}} \\[0.8ex]
\cline{4-6} \cline{7-9}
& & &
95\% CI &
\begin{tabular}[c]{@{}c@{}}Target within \\ posterior\end{tabular} &
\begin{tabular}[c]{@{}c@{}}MAP distance \\ from target\end{tabular} &
95\% CI &
\begin{tabular}[c]{@{}c@{}}Target within \\ posterior\end{tabular} &
\begin{tabular}[c]{@{}c@{}}MAP distance \\ from target\end{tabular} \\
\hline
$\alpha_{anterior}$ & - & 2.12 & [1.42, 3.62] & $\checkmark$ & 0.33 & [0.77, 3.83] & $\checkmark$ & 0.42\\
$\alpha_{posterior}$ & - & 1.42 & [0.73, 3.42] & $\checkmark$ & 0.11 & [0.34, 3.83] & $\checkmark$ & 0.01\\
$\alpha_{septum}$ & - & 2.57 & [1.16, 3.75] & $\checkmark$ & 0.35 & [0.54, 3.91] & $\checkmark$ & 0.28 \\
$\alpha_{lateral}$ & - & 2.71 & [1.21, 3.72] & $\checkmark$ & 0.03 & [0.53, 3.87] & $\checkmark$ & 0.53\\
$\alpha_{roof}$ & - & 2.78 & [1.12, 3.47] & $\checkmark$ & 1.25 & [0.67, 3.84] & $\checkmark$ & 1.22\\
EDP & mmHg & 4.50 & [3.17, 6.38] & $\checkmark$ & 0.38 & [2.18, 10.19] & $\checkmark$ & 0.38 \\
ESP & mmHg & 29.6 & [20.1, 34.9] & $\checkmark$ & 1.05 & [14.6, 36.8] & $\checkmark$ & 1.50 \\
$k_{peri}$ & kPa/$\mu$m & 0.003 & [0.001,0.004] & $\checkmark$ & 0.0004 & [0.0003, 0.004] & $\checkmark$ & 0.0004 \\
PTH & - & 0.60 & [0.47, 0.85] & $\checkmark$ & 0.01 & [0.46, 0.92] & $\checkmark$ & 0.12\\
\hline
\end{tabular}%
} 
\end{adjustwidth}
\end{table}

\subsection*{Effect of anatomical and myocardial stiffness heterogeneity on left atrial deformation}

To confirm the importance of regional stiffness and limited impact of heterogeneous wall thickness on regional displacements, we examined the differences in the LA model function when a uniform thickness and a global material model were used. 

\begin{figure}[H]
    \centering
    \includegraphics[width=\linewidth]{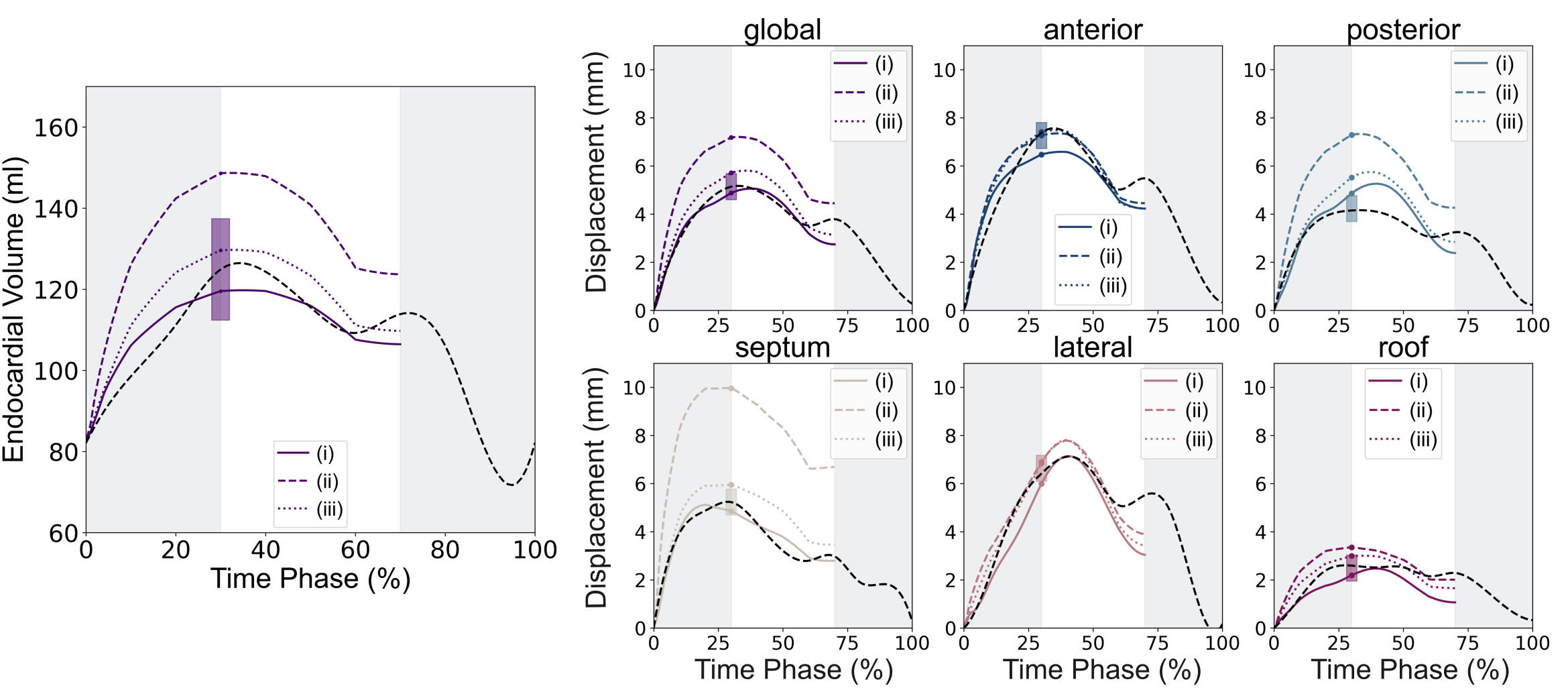}
    \caption{\textbf{Comparison of model function}. The panels show the global and regional displacements as well as the volumes obtained when the simulator was evaluated using the MAP estimate derived from three model set-ups: (i) model includes patient-specific LA wall thickness and regionally varying stiffness parameters; (ii) model includes patient-specific LA wall thickness and a single global stiffness parameter; (iii) model includes uniform LA wall thickness and regionally varying stiffness parameters. The dashed black line indicates the LA behaviour derived from the CT image set and the shaded box represents  95\% confidence interval on the target value.} 
    \label{fig:comparing_models}
\end{figure}

The plots in Fig \ref{fig:comparing_models} shows that similar calibration performance was obtained when regional stiffness parameters were used, irrespective of the inclusion of a patient-specific wall thickness. However, when the model included only a single global stiffness metric, our ability to recover both the global and regional image-derived behaviour is reduced as shown in Fig \ref{fig:comparing_models}. Across the cohort, the average RMSE between the simulated and CT-derived displacements at ES was $0.97 \pm 0.36$ mm when a non patient-specific wall thickness was used vs. 0.90 $\pm$ 0.39 mm when the patient-specific wall thickness was used in the model. When regional heterogeneity in myocardial stiffness was replaced by a globally applied stiffness, this average error increased to 1.43 $\pm$ 0.45 mm, with greater discrepancies seen among the regional displacements (see Supporting Information \nameref{S12_File}). These results further support our previous findings that regional variations in wall thickness do not play a significant role in dictating regional heterogeneity in passive LA biomechanics. Additionally, this provides further evidence towards our hypothesis that LA myocardial stiffness varies regionally and contributes significantly to local LA mechanics.

\section*{Discussion}

In this study, we investigated the contribution of regional myocardial stiffness and anatomy to image-derived estimates of regional LA biomechanics using a cohort of 10 patient-specific LA models from HF patients. We have shown that regional wall thickness and regional EAT do not significantly contribute to heterogeneous LA displacement patterns during the reservoir phase, while regional stiffness plays a relevant role.  Through this, we have demonstrated the feasibility of personalised regional stiffness estimation in the atria which has not been attempted previously.

\added{To determine regional stiffness estimates, we have applied HM, an approach that first searches for a region of physiologically non-implausible parameter sets within an explicitly defined parameter space. We performed 200 simulations in the first wave of HM, followed by 100 simulations in each subsequent wave and conducted 100 000 emulator evaluations at each wave, over an average of 4 waves. With this simulated and emulated data, we identified implausible and non-implausible regions of the input space on a patient-specific basis. We then used this physiologically non-implausible space, as determined by HM, as a prior to estimate the posterior distribution of the input parameters using MCMC. Through this framework, we did not define a singular unique parameter set that describes the clinical deformation data for each patient. Instead, we estimated the posterior distribution of the input parameters, with the width of the posterior distributions providing a measure of the uncertainty of the parameter prediction we obtained. We then validated our model performance against patient-specific imaging data.} 

 In doing this, we showed that in HF patients, myocardial stiffness varies across LA regions. However, we did not find any region to be consistently stiffer across the cohort and our findings revealed that stiffer regions were not associated with increased fat volume or increased wall thickness.

\subsection*{Clinical Data and Feature Tracking Accuracy}

This work describes a methodology for regional parameter estimation in the LA. Previously, other studies aimed at estimating myocardial stiffness have mainly focused on ventricular myocardium and considered only a global quantification of stiffness \cite{Ghista1980CardiacStates, Augenstein2005MethodImaging, Wang2009ModellingFunction, Xi2011AnEstimation, Genet2014DistributionTreatments, Wang2018LeftAnalysis, Palit2018InMyocardium, Marx2022RobustConfiguration}. Attempts at performing parameter estimation in the atria were previously impacted by limitations in available imaging modalities to adequately capture the thin-walled atria. The resolution of cardiac MR images \cite{Wang2018LeftAnalysis, Xi2013TheMeasurements} is typically of the same, or greater, order of magnitude as LA wall thickness \cite{Ho2012LeftRevisited} and while the high resolution of CT imaging is capable of capturing atrial anatomy, the high-radiation dose associated with the imaging limits its clinical indication. In this study, retrospective gated contrast-enhanced CT with an in-plane isotropic resolution between 0.32 and 0.49 mm and slice thickness between 0.4 and 1.0 mm was used. These images provide high-fidelity images of the LA anatomy thus, aiding in accurate 3D feature tracking and minimising the uncertainty associated with the key image-derived target features. The feature tracking implemented in this study was previously verified in \cite{Sillett2024AFibrillation} where the authors found a median RMSE of less than 0.8 mm in LA endocardial surface coordinates at ES across a 30 patient cohort, suggesting that our feature tracking methods accurately estimate passive LA deformation and provide acceptable targets for calibration.

\subsection*{Calibration Methods}

In this study, we used a three-step approach to infer regional stiffness parameters, involving the selection of the most relevant input parameters using GSA, constraining the input space using HM and finally identification of a calibrated input parameter set.

\deleted{While limited examples exist of using strain or displacements to estimate atrial stiffness, ventricular studies predominantly employ displacement-based cost functions. A comparison of strain-based and displacement-based cost functions for parameterisation using the Guccione material law showed similar performance when either Green–Lagrange strain or displacement was used as the metric of deformation.}
\deleted{The complex geometry of the atria means that both deformation or strain in one region can influence the strain and stiffness in another. For instance, strain in one area can induce shape or curvature changes in a distant region. This, in turn, affects how local loads impact that region and, consequently, the inferred stiffness. Therefore, while both strain and deformation-based cost functions present challenges, deformation offers advantages. It is simpler to calculate, less susceptible to noise as it avoids gradient calculations, and results in simulations that can be visually compared with medical images. Thus, in this study, we chose to use displacement as our main metric of LA deformation.}

\added{In this study, we employed a displacement-based cost function, without considering any alternatives. This choice was informed by previous studies, which predominantly employed displacement-based cost functions \cite{Ghista1980CardiacStates, Wang2009ModellingFunction, Xi2011MyocardialFilter, Xi2011AnEstimation, Genet2014DistributionTreatments}. However, energy and strain-based cost functions may also be used, though limited examples following a strain-based approach exist \cite{Nasopoulou2017ImprovedFunction}. These alternative cost functions have specific merits, with potentially the most important benefit over a displacement-based metric being that a strain-based metric would not be affected by rigid body motion and thus, it may potentially be more indicative of local material information. However, there are multiple strain indexes that may be considered, such as principal strains, fibre strains, Green–Lagrange strains or strain invariants, and the most suitable choice of index for this type of analysis is not clear. Further, strain-based cost functions require calculation of gradients, which can increase noise \cite{Hild2006DigitalReview}. There are trade-offs associated with both the choice of strain and displacement cost functions and, as evidenced by the different cost function choices made in the myocardium, the optimal cost function for passive stiffness estimates remains an open area of research. Therefore, while there are definite benefits to a strain-based approach for stiffness characteristation, we chose to use a displacement-based cost function as this followed an approach which has been shown to be successful on many occasions \cite{Wang2009ModellingFunction, Xi2011MyocardialFilter, Xi2011AnEstimation, Genet2014DistributionTreatments}, with the only previous attempt at regional stiffness estimation also employing a displacement-based method \cite{Ghista1980CardiacStates}.}

Through GSA, we showed that in the Guccione material law (eq \ref{eq:Guccione_Q_reformulated}), the $\alpha$ scaling parameter consistently plays a more relevant role in influencing deformation in the LA and as such, it was the parameter we chose to fit. Although the physiological meaning of $C$ and $\alpha$ is similar and some previous studies have chosen to fit only $C$ \cite{Wang2018LeftAnalysis}, $C$ scales the strain energy function linearly, while the effects of $\alpha$ are exponential. This fact could justify why small changes in $\alpha$ have greater impact in changes in volume and displacement compared to $C$, and therefore, it is more relevant to calibrate it accurately.

Here, we used HM as a pre-calibration space reduction technique followed by MCMC for parameter calibration. 

To calibrate our model, we fit to CT volume measurements as well as global and regional displacements derived from the patient image data. Using both global and regional measurements derived from the data ensured that local characteristics of LA deformation were preserved in our simulated model. \added{In our study, we fitted the model parameters to the deformation at the ES frame. This frame exhibited the largest deformation, likely providing the strongest signal for model fitting. While previous models have used time-resolved CT image data to personalise material parameters for the LA myocardium and achieved relative errors in LA cavity volume between image and simulated data of less than 1\% \cite{Shi2024AnMechanics}, Fig \ref{fig:calibrated_outputs} demonstrates that fitting regional deformations solely to the ES frame still recovers the transient deformation trace, which may limit the benefit of incorporating additional frames. We used the volumes and deformations from the time-resolved imaging data as a means of validation for our calibration. In doing so, we facilitate the use of our framework in cases where only limited clinical imaging data is available.}

The sequential HM approach implemented was an important precursor for calibration using MCMC, not only because it reduces the size of the non-implausible region to be explored, but because the accuracy of the GPEs improve with each successive wave due to decreasing emulator uncertainty as more training samples are added to the area of interest. 

Both HM and MCMC allowed us to fit 9 input parameters concurrently. This avoids the issues created by sequential parameter fitting, as used in \cite{Wang2009ModellingFunction, Augenstein2005MethodImaging}. In sequential parameter fitting, the final parameter values may be dependent on the initial choice of values and error tends to accumulate in the latter fitted parameters, while in concurrent fitting, error is evenly distributed across parameters. As mentioned previously, HM does not ensure that individual input parameter ranges are reduced but rather focuses on a whole input space reduction as shown in Fig \ref{fig:HM}. In previous studies, through this iterative space reduction, the parameter ranges are reduced \cite{Coveney2018FittingMatching, Longobardi2020PredictingRats}. Across our cohort, HM did reduce the input space by 45\% to 99\%, however, due to parameter coupling, it did not always reduce the bounds on individual parameter ranges. 

As illustrated in \cite{Coveney2018FittingMatching}, the HM plots in Fig \ref{fig:HM} reveal pairs of input that might co-vary to match the target output features. Most consistently over our patient cohort, the HM plots converged towards a coupled EDP - ESP parameter space as observed in 7 out of 10 cases. The cornerplots produced by MCMC (Fig \ref{fig:cornerplot}) also reveal the coupled relationship between EDP and ESP, shown by the diagonal contour plots in the EDP - ESP space. This suggests that non-unique combinations of EDP and ESP exist that could potentially satisfy the observed LA behaviour. To address this issue of non-uniqueness directly, clinically obtained simultaneous pressure recordings to accompany the gated CT image set would be necessary. Cardiac catheterisation provides a direct measure of chamber pressure, however it is an invasive technique, thus limiting its application. Estimates of pressure from Doppler echocardiography provide a non-invasive alternative to this approach \cite{Tachjian2019EstimationStudy}, but the accuracy of the non-invasive approach can be limited compared to the invasive measurement and can vary considerably among patient groups \cite{Bowcock2022BedsideGem}.

\deleted{The use of MCMC for calibration helps us to deal with the issue of parameter coupling. Previous studies have utilised deterministic methods for parameter inference including genetic algorithms and sequential quadratic programming. Xi et al. used a reduced order unscented Kalman filter for parameter estimation that incorporated noisy measurements of image-derived deformation and showed how the target uncertainty propagated to the error associated with the parameter estimates. MCMC is a stochastic approach that approximates the posterior distribution of the input space and provides a set of values that are most likely to return the target features instead of a single estimate. The width of the distribution provides a measure of the uncertainty of the parameter prediction we obtain in this work }

\added{Generally, multi-scale physiological models are complex and often include a large number of parameters, so coupling between input parameters, as seen in EDP - ESP space, is not uncommon. This, combined with inter-patient variability and noise associated with clinical data, makes parameter calibration in these models a well-recognised challenging task \cite{Argus2022AutomatedModel}. Previous studies have attempted to perform parameter inference using deterministic methods including genetic algorithms \cite{Palit2018InMyocardium} and sequential quadratic programming \cite{Augenstein2005MethodImaging, Wang2009ModellingFunction}. However, the nonlinearity in the relationship between simulated mechanical behavior and material parameters and the interdependence among these material parameters, typically impairs the ability of the optimizer to accurately estimate parameter values. This issue is further compounded by the presence of measurement noise as demonstrated in \cite{Xi2011MyocardialFilter}.}

\added{By using Bayesian methods, we account for uncertainty in both the model parameters and the calibration data \cite{Willmann2022BayesianDeformation}. Bayesian methods do not provide a single unique solution but instead, the posterior distribution returns the probability density for all input parameters \cite{Willmann2022BayesianDeformation}. Here, we used MCMC, a Bayesian approach that approximated the posterior distribution of the input parameters within the prescribed physiological parameter ranges. Following a Bayesian approach allowed us to encode the knowledge gained from HM about implausible and non-implausible parameter values within the prior distribution. The width of the posterior distribution provided a measure of the uncertainty of the parameter prediction we obtain in this work. In Supporting Material S10, we demonstrate how we used MCMC to obtain posterior distributions for all input parameters that recapitulate the image-derived output features. From these distributions, we gained information about the relative differences in stiffness across regions, informing LA stiffness patterns.}

\subsection*{The effect of anatomy}

Despite the heterogeneous wall thickness of the LA, it is often modelled with a uniform wall thickness \cite{Moyer2015ChangesModel,Adeniran2015EffectsAtria,Phung2017EffectAtrium} due to challenges in capturing the thin-walled structure with available imaging techniques. The high resolution of CT imaging allowed us to capture patient-specific wall thickness in the LA and evaluate its contribution to regional LA biomechanics. However, clinically, indication for retrospective gated CT is limited due to the high level of X-ray exposure to patients. Our findings suggest that regional wall thickness is not correlated to regional reservoir displacement. We provided a direct comparison of calibration results when the LA is modelled with a patient-specific wall thickness vs a uniform wall thickness in the Supporting Information \nameref{S12_File}. Using the uniform wall thickness model, we fit the global and regional displacements and volume with a mean RMSE of $0.97 \pm 0.36$ mm and  $\pm 11.3$ ml over the cohort (vs mean RMSE of $0.90 \pm 0.39$ mm and $\pm 11.0$ ml using the patient-specific wall thickness model). The agreement between the calibration results supports the use of a uniform wall thickness for the LA when investigating passive LA mechanical behaviour. Additionally, from our mixed effect model, regional myocardial stiffness does not depend significantly on regional wall thickness. The HM results obtained using a uniform wall thickness could not conclude a different set of stiffness parameter ranges from those obtained when the model included a patient-specific wall thickness. 

\replaced{For our cohort of patients, these results suggest that wall thickness did not influence passive LA biomechanics as explored in this study. However, previous studies not accounting for the effect of the pericardium or incorporating the physiological motion of the MV \cite{Feng2019AnalysisValve,Augustin2020TheAtrium} found a link between wall thickness and deformation.} {From this, LA wall thickness does not appear to an important determinant of LA biomechanics.} In Feng \cite{Feng2019AnalysisValve}, simulations were performed using a patient-derived wall thickness and a uniform wall thickness. Using a uniform wall thickness led to a reduction in average fibre strain and principal stress measured on the atrial wall compared to using a non-uniform thickness \cite{Feng2019AnalysisValve}. Augustin et al. looked at the correlation of simulated LA wall stress with wall thickness and curvature using three patient cases \cite{Augustin2020TheAtrium}. They found an average Spearman's correlation value of 0.598 between inverse LA wall thickness and wall stress, indicating some dependency of LA wall stress on wall thickness. In these studies, the uniform stiffness or the exclusion of some physiological boundary conditions, including the pericardium and the pulling action of the left ventricle, may have contributed to the discrepancies in the associations observed. The results of the GSA show that pericardium stiffness ($k_{peri}$) and the pericardium penalty threshold (PTH) sometimes appear among the top ranked parameters for sensitivity. No consideration of these relevant boundary conditions may lead to greater dependence of LA deformation on wall thickness. \added{To demonstrate that wall thickness does not have to be a strong predictor of biomechanics, in Supporting Information \nameref{S13_File}, we show, in a simplified geometry, that the association between wall thickness and deformation is dependent on anatomy, the locations of thicker regions and the physiological and boundary constraints. }

Augustin et al. used the law of Laplace as a model for wall stress in the LA during deformation \cite{Augustin2020TheAtrium}. Following this approach, we also used a mixed effect model to examine the relationship between regional stiffness and regional strain and wall thickness and found no significant correlation (see Supporting Information \nameref{S14_File}). The constraints on LA deformation imposed by the pulmonary veins, MV and pericardium are not accounted for in the law of Laplace and therefore may make the Laplace law unsuitable for use in the LA. This aligns with the conclusion of the authors in Augustin et al. \cite{Augustin2020TheAtrium}.

\subsection*{Spatial heterogeneity in myocardial stiffness}

Currently, in modelling myocardium, uniform material properties are often assumed in the ventricles and in the atria \cite{Adeniran2015EffectsAtria, Feng2019AnalysisValve, Augustin2020TheAtrium}. The assumption often remains when modelling both healthy and pathological scenarios. The presence fibrosis in the LA has been represented using the Guccione material law by increasing the isotropic stiffness term \cite{Moyer2015ChangesModel} however, this increased stiffness was applied uniformly across the LA, without any regional considerations \cite{Benito2018PreferentialFibrillation, Assaf2023EffectII}.

Using a global stiffness parameter, our simulations were able to capture some of the LA deformations estimated from retrospective gated CT but could not reproduce all the regional displacements. This supports the hypothesis that myocardial stiffness varies across the LA and suggests that global metrics such as volume or global displacement might not provide enough information to reveal regional heterogeneities in material parameters when used for calibration purposes. Moyer et al. \cite{Moyer2015ChangesModel} presented an LA mechanics model that used a uniformly applied transversely isotropic material law to represent LA myocardium. This study considered regional motion in the validation of their LA model. However, this study used average values of regional emptying fraction taken from a cohort patients rather than fitting to patient-specific values \cite{Moyer2015ChangesModel}. Our results showed that regional differences in stiffness within an individual did not have a common pattern over the cohort. This might suggest that fitting stiffness parameters to the population reduces the need for regional fitting. 

We found that heterogeneous stiffness across regions of the LA is necessary to capture physiological regional mechanical behaviour in HF patients. This study is the first to highlight regional differences in myocardial stiffness across the LA and suggest that, in the future, the inclusion of spatial heterogeneities in cardiac models in myocardial stiffness might be considered if local LA biomechanics are of interest. The regional differences in stiffness play a role in dictating observed LA function and may potentially provide an indication of disease. The presence of fibrosis might be an underlying contributor to changes in myocardial stiffness. It has been shown that fibrosis as quantified by LGE-MRI was not uniformly distributed across the LA wall \cite{Benito2018PreferentialFibrillation, Assaf2023EffectII}. One study by Benito et al. showed, in a 62 patient cohort, that the roof and posterior wall tended to be more affected by fibrosis as identified by LGE-MRI \cite{Benito2018PreferentialFibrillation}. This heterogeneous distribution of fibrosis may be responsible for the variation in patient myocardial stiffness observed in our cohort.

Fibrosis is likely to be present in numerous cardiac conditions but may also exists in healthy individuals. Fibrosis contributes to the onset of AF \cite{Habibi2015AssociationAF, Platonov2011StructuralAge} and is a key factor in the maintenance and progression of the arrhythmia \cite{Kuppahally2010LeftFibrillation, Marrouche2014AssociationStudy}. In the context of AF, the fibrosis burden present in the atria is of particular interest for patient stratification \cite{Marrouche2014AssociationStudy}, predicting the likelihood of AF recurrence \cite{McGann2014AtrialMRI} and deciding on optimal ablation strategies \cite{Sim2019LeftSubstrate}. The presence of fibrosis is expected to alter myocardial material properties, resulting in increased stiffness. Current methods for identifying fibrosis in the LA include voltage mapping \cite {Sim2019LeftSubstrate} and LGE-MRI \cite{Siebermair2017AssessmentImplications}. However, there are limitations associated with both methods \cite{McGann2014AtrialMRI, Sim2019LeftSubstrate} and controversy regarding agreement between the two methods \cite{Malcolme-Lawes2013AutomatedStudy, Jadidi2013InverseMapping}. The ability to approximate fibrosis burden through non-invasively obtained atrial biomechanics provides a viable alternative to existing techniques for fibrosis quantification. Thus, an extension of this work might involve obtaining increased numbers of regional patient-specific stiffness values, using our calibration pipeline, along with regional estimations of fibrosis from voltage mapping or LGE-MRI. This would allow for investigation into a link between myocardial stiffness and the presence of fibrosis.

\subsection*{Limitations}

\deleted{In this study, we have used a small cohort of patients imaged at a single centre. Future work should look to expand to larger cohort using data from multiple centres.}

\added{In this study, we aimed to perform parameter inference by matching simulated deformations to those observed from images. Thus, we used image-derived MV displacements to drive deformation in our simulation model for all parameter combinations. Physiologically, MV annulus displacements would vary with different myocardial stiffnesses and pressures. However, for the context of this study, it was relevant to maintain image-based characteristics like anatomy and MV displacement to facilitate matching of simulated LA motion to the imaging data. For applications outside of model calibration, and outside the scope of this study, the incorporation of dynamic coupling or pressure-driven boundary conditions at the MV annulus may increase the predictive capacity of the model.}

\added{We used RMSE between our image-derived and simulated deformations at ES to assess calibration quality. Previous studies have commonly accepted RMSE thresholds between 1 to 2 mm as indicative of acceptable fitting performance \cite{Wang2009ModellingFunction,Xi2013TheMeasurements, Nasopoulou2017ImprovedFunction}. The calibration framework developed and employed in this study achieved good overall performance, with an average RMSE of less than 1 mm ($0.90 \pm 0.39$) across the cohort. This aligns well with the uncertainty associated with the image-derived deformation estimates (median RMSE of 0.8 mm) \cite{Sillett2024AFibrillation}. However, in some cases, larger variations in the agreement between simulated and image-derived deformations were observed. Here, we highlight some of the limitations that may be responsible for these discrepancies.}

 We used a volume-based convergence criteria to find the unloaded reference configuration \cite{Marx2022RobustConfiguration}. With this approach, the simulated ED configuration would match image ED configuration within a tolerance of 1.0\% of the target ED volume. Though ED volumes between the simulated and image configurations matched, the complex shape of the LA lead to shape discrepancies between the simulated and target ED configurations in some cases. This can result in a poorer fit ($> 2.0$ mm) as we saw in some cases in Supporting Information \nameref{S10_File}. \replaced{This occurred in only 16\% of the regional fitted displacements with 75\% of regions having a maximum error of $0.90$ mm.}{This occurred in only 16\% of the regional fitted displacements with the anterior region being affected 50\% of the time when the shape discrepancy occurred.}

\deleted{The computational load associated with this calibration pipeline is large. For each case, we ran 200 simulations in the first wave of HM, followed by 100 simulations in each subsequent wave over an average 4 waves, at a cost of approximately 53 cpu-hours per simulation. Across the cohort, this totals to over 250 000 core-hours. The use of GPEs is important to reduce the computational expense but many expensive runs of the simulator are still necessary to train the emulators through successive waves of HM. As such, the computational cost remains high. The high computational cost limits the feasibility of this pipeline being applied in clinical settings.}

\deleted{We also did not have access to simultaneous pressure recordings to accompany the gated CT image set. Thus, pressure values at ventricular ED and ES were included as parameters to fit in our model. From the GSA results, the EDP and ESP had high sensitivity on volume and displacement at ES. Therefore, the inclusion of patient-specific pressure data would reduce model uncertainty and would decrease the number of parameters to be inferred, which in turn reduces computational expense as less simulation data is needed to train the GPEs.}

 In MCMC, the choice of the prior distribution may play a role in accurate parameter estimation \cite{Lambert2005HowWinBUGS, Argus2022AutomatedModel}. Here, we chose to use uniform prior distribution for all input parameters, defined within the parameter bounds of the NROY space from the final wave of HM. This approach was chosen as there was no evidence to bias the prior distribution towards any specific value\added{.} \deleted{and it is simple and easy to implement.} However, it is possible that the choice of a naive prior might have an influence on the inferences made using MCMC \replaced{and}{Though in this application, the use of a uniform prior did not appear to create a problem, for more complex applications,} the choice of a more informed prior might improve convergence speed and accuracy \cite{Lambert2005HowWinBUGS}. 

 \added{Within the framework presented in this study, we attempted to infer atrial pressures at ED and ES. The sensitivity analysis revealed that across our cohort, atrial pressures were important determinants of volume and regional displacements at ES. However, we did not have access to pressure measurements in these patients. To account for this, we introduced a physiologically non-implausible prior for pressure. The lack of pressure measurements means that we have higher uncertainty in our regional stiffness posterior distributions. But, our framework allows us to estimate regional stiffness while accounting for uncertainty introduced using the data we have available. Further, in our verification study we were able to show that, even without known pressures, we were able to recover regional variations in stiffness.}

\added{Here, we also highlight nine potential sources of model mismatch, that is, discrepancy between the mathematical model employed and the physiological system \cite{Paun2020AssessingCirculation}.} \deleted{ In our study, we did not account for model mismatch, that is, discrepancy between the mathematical model employed and the physiological system. Here, we highlight three potential sources of model mismatch.} 

(i) We modelled the LA in isolation. We incorporated influence of surrounding structures by inclusion of the boundary conditions, namely the pericardium and the image-derived MV displacements. However, the use of four-chamber models for patient-specific modelling of cardiac mechanics is becoming more widespread \cite{Strocchi2023CellEmulators}. Performing calibration studies with such models, though more complex, might more closely mimic physiological conditions and thus improve calibration results. \added{However, four-chamber models would require a significantly higher computational cost compared to our LA model. The development of more efficient software and HPC facilities would facilitate the extension of this approach to the whole heart in future.} 

\added{(ii) The CT imaging sequence used to construct the LA anatomical models was not optimised to cover the left atrial appendage as such, in some cases, the images only partially covered the appendage. For consistency, the appendage was excluded from this analysis. However, in this study, we aimed to estimate regional stiffness parameters in the LA body only. Therefore, the inclusion of the appendage was not likely to affect the study findings.}

\added{(iii) We quantified LA EAT from CT but did not include it as a mechanical layer in our simulations. Previous biomechanical models have modelled EAT as its own subdomain within the simulation framework \cite{Pfaller2019TheModeling} and integrated cardiac and EAT interaction by adjusting pericardial spring stiffness so that the presence of EAT leaves those areas more free to move \cite{Fedele2023AElectromechanics}. The physical presence of EAT on the LA myocardium during simulation may affect its deformation and, subsequently, model calibration. However, from our image-based analysis, we did not see a significant correlation between regional EAT volume and regional deformation. Thus, we chose to exclude EAT from our simulations and avoid the additional model complexity its inclusion would introduce.}

\added{(iv) Assumptions were also made surrounding the myocardial material. We used the transversely isotropic Guccione material law for atrial myocardium, which does not fully describe the orthotropic nature of atrial myocardium. In the Guccione model, we fixed the anisotropy ratio between the anisotropic stiffness parameters (eq. \ref{eq:Guccione_Q_reformulated}). This simplification was taken from the literature, where it is commonly done to reduce model complexity and facilitate the regional parameter estimation procedure \cite{Nasopoulou2017ImprovedFunction}.}

\replaced{(v)}{(ii)} We used a rule-based fibre field as opposed to patient-specific fibres in our simulations. In-vivo assessment of myofibril orientation, while increasingly feasible, remains outside the scope of standard clinical practice, particularly in the atria. Thus, our ability to use patient-specific fibre architectures is limited. \added{As such, fibre architectures generated using rule-based methods as well as Laplace-Dirichlet rule-based methods, such as that proposed by Piersanti et al., are commonly used and typically demonstrate physiological concordance \cite{Piersanti2025DefiningTwins}.}

\added{(vi) The assumptions and simplifications made in the choice of boundary conditions used in this model should also be noted. A homogeneous pressure load is assumed to act on the endocardial boundary during passive LA filling and emptying. However, variation in endocardial pressure from pulmonary veins to MV annulus is likely and this, in turn, could impact the local deformation. Using a homogeneous pressure load is a commonly accepted simplification \cite{Adeniran2015EffectsAtria,Moyer2015ChangesModel,Augustin2020TheAtrium}, supported by reported cavity pressure variations \cite{Bowcock2022BedsideGem}.}

\added{(vii) To estimate the stress-free configuration, we run an unloading simulation. To recover the image-derived MV motion would require either a model incorporating MV ring stiffness (similar to the approach followed in the atrial body) or defining boundary conditions on the MV ring. As we do not know, nor are we estimating, the stiffness of the MV, we have fixed the MV ring to its location at ED during unloading and applied image-derived time-varying boundary conditions during simulation. ED is defined as the CT image during the peak of the R-wave. At this point, the MV is closed and the atria are likely relaxed, as otherwise ventricular contraction can reactivate atrial contraction \cite{Land2018InfluenceFunction}. The MV ring location can be precisely defined from the CT images and fixing its location during unloading ensures that the reinflated atrium matches the desired ED anatomy. However, this is an approximation. As such, it may not reflect a good estimate of the stress-free MV ring, and may impact the estimation of the atrial myocardium stress-free configuration. However, by fixing the MV annulus during the unloading step, we ensured a consistent reference configuration for applying image-derived mitral annular displacements during simulation.}

\added{(viii) Due to the lack of availability of patient-specific LA pressure data, we used an idealised pressure transient to drive deformation in our simulated model. This idealised transient reflects a simplification of true LA pressure dynamics. However, one key feature of the idealised pressure transient was that the timing of the peak in LA pressure matched that of the peak in LA volume determined from the image data, occurring at the ES time-point. As we were most interested in matching patient LA deformation at ES with our simulated model, this ensured that pressure behaviour at the point of interest was in accordance with physiological behaviour.}

\replaced{(ix)}{(iii)} LA function comprises a passive filling and emptying component followed by an active booster pump component. Here, we only modelled the passive behaviour of the LA. For the active phases, electrical activation of the LA would be needed. This would affect the motion of the MV, and therefore a different simulation setup would be needed. \added{However, for the task of stiffness estimation, the passive filling function of the LA provided the necessary model behaviour.}

\added{Across this cohort and in the general population, variations in LA size and shape can be observed. Thus, it is common to use body surface area normalisation when comparing LA metrics among patients \cite{Jeyaprakash2021APopulation}. The clinical dataset available to us did not include any information on patient body surface area, height or weight. Therefore, we were not able to perform such normalisation during our analysis.}

\added{Finally, the computational load associated with this calibration pipeline is large. For each case, we ran 200 simulations in the first wave of HM, followed by 100 simulations in each subsequent wave over an average 4 waves, at a cost of approximately 53 cpu-hours per simulation. Across the cohort, this totals to over 250 000 core-hours. The use of GPEs is important to reduce the computational expense but many expensive runs of the simulator are still necessary to train the emulators through successive waves of HM. As such, the computational cost remains high. The high computational cost limits the feasibility of this pipeline being applied in clinical settings. }

\section*{Conclusion}

We constructed a modelling framework that was able to replicate passive global and regional deformation in the LA of 10 HF patients. With this framework, we performed the first regional stiffness estimation in the LA. Our results suggest that heterogeneity in myocardial stiffness plays a role in dictating global and regional LA deformation in HF patients while regional variations in anatomy including wall thickness and EAT, are less important. Underlying factors that might dictate this regional variation in stiffness, for example the presence of fibrosis, requires further investigation. Nevertheless, our study showcases the use of computational modelling to estimate tissue properties which remain elusive to in-vivo clinical measurements.

\section*{Acknowledgments}

This work used the ARCHER2 UK National Supercomputing Service (https://www.archer2.ac.uk).

\nolinenumbers

%
%
%


\section*{Supporting Information}

\paragraph*{\added{S1 file}.} \label{S1_File} 
\added{\textbf{Model generation from CT images.}} \added{Description showing how LA blood pool, myocardium and EAT were segmented from patient images.}

\paragraph*{\replaced{S2}{S1} file.} \label{S2_File} 
\textbf{Sensitivity of EAT quantification.} We compared the effects of EAT volume on atrial biomechanics using different methods of EAT quantification.

\paragraph*{\replaced{S3}{S2} file.} \label{S3_File} 
\textbf{Mesh convergence.} We compared model results using a high resolution, high computational resource mesh and a lower resolution, less computationally demanding mesh.

\paragraph*{\replaced{S4}{S3} file.} \label{S4_File} 
\textbf{Region definition.} Detailed explanation of how the five regions on the LA were defined using UACs.

\paragraph*{\replaced{S5}{S4} file.} \label{S5_File} 
\textbf{Wall thickness calculation.} We used the eikonal equation to obtain patient-specific LA wall thickness maps. We compared the performance of this fast-evaluating method to a slower, more computationally expensive but previously validated approach.

\paragraph*{\replaced{S6}{S5} file.} \label{S6_File} 
\textbf{Pericardium penalty map.} Detailed explanation of the pericardium definition.

\paragraph*{\replaced{S7}{S6} file.} \label{S7_File} 
\textbf{ Passive mechanics sensitivity analysis.} We used GSA to identify the most relevant regional stiffness parameters in each case of the 10-patient cohort.

\paragraph*{\replaced{S8}{S7} file.} \label{S8_File} 
\textbf{Gaussian process emulation.} Summary of the GPE training results for all patient cases in this study.


\paragraph*{\replaced{S9}{S10} file.} \label{S9_File} 
\textbf{History Matching.} Summary of the convergence of the NROY space for all patient cases in this study.


\paragraph*{\replaced{S10}{S11} file.} \label{S10_File} 
\textbf{Calibration results using MCMC.} Summary of the final calibration results using MCMC for each of the 10 patient cases.

\paragraph*{\replaced{S11}{S12} file.} \label{S11_File} 
\textbf{Model Validation.} We calibrated our model to CT-derived deformations at ES. We validated the model against the global and regional deformation transients during the reservoir and conduit phases.

\paragraph*{\replaced{S12}{S13} file.} \label{S12_File} 
\textbf{Calibration performance.} Comparison of the calibrated model performance to the baseline model when regional heterogeneity in myocardial stiffness is not considered and when a uniform LA wall thickness is assumed.

\paragraph*{\added{S13} file.} \label{S13_File}
\added{\textbf{Association between wall thickness and deformation in simplified geometry.}} \added{We used a hemispherical geometry and simplified boundary conditions to examine the relationship between wall thickness and deformation.}

\paragraph*{\replaced{S14}{S15} file.} \label{S14_File} 
\textbf{Consideration of the law of Laplace.} We examined the suitability of the law of Laplace for application to the LA.

\end{document}